\documentclass[11pt]{article}
\usepackage{graphicx} 
\usepackage{xcolor}
\usepackage[numbers,round,semicolon]{natbib}
\usepackage{amsmath}
\usepackage{amssymb}
\usepackage{bbm}
\usepackage[figuresright]{rotating}
\usepackage{multicol,ragged2e}
\usepackage{setspace}
\usepackage{algorithm}
\usepackage{algpseudocode}
\usepackage{authblk} 
\usepackage{subcaption}
\usepackage{hhline}
\usepackage{url}

\newtheorem{assumption}{Assumption}
\newtheorem{identres}{Identification Result}

 \newcounter{acounter}
\title{Estimating treatment effects with a unified semi-parametric difference-in-differences approach}

\author[1]{Julia C. Thome}
\author[1]{Andrew J. Spieker}
\author[1]{Peter F. Rebeiro}
\author[2]{Chun Li}
\author[3]{Tong Li}
\author[1,*]{Bryan E. Shepherd}

\affil[1]{Department of Biostatistics, Vanderbilt University, Nashville, Tennessee, U.S.A.}
\affil[2]{Department of Population and Public Health Sciences, University of Southern California, Los Angeles, California, U.S.A}
\affil[3]{Department of Economics, Vanderbilt University, Nashville, Tennessee, U.S.A.}
\affil[*]{Corresponding author: bryan.shepherd@vanderbilt.edu}
\date{}

\begin{document}

\maketitle
\begin{abstract}
Difference-in-differences (DID) approaches estimate causal effects with observational data before and after an intervention. DID traditionally estimates the average treatment effect among the treated and assumes parallel trends in outcome means. With skewed outcomes, a transformation is often needed to compare means; however, transformations may be difficult to choose and results may be sensitive to the choice and must be interpreted on the transformed scale. Recent DID methods estimate alternative treatment effects that offer additional insights. However, each estimator requires a different parallel trends assumption. We introduce a new DID method that estimates average, quantile, probability, and novel Mann-Whitney treatment effects among the treated with a single unifying parallel trends assumption. The method uses a semi-parametric cumulative probability model (CPM): a linear model for a latent variable on covariates, where the latent variable results from an unspecified transformation of the outcome. Our approach makes a single parallel trends assumption on the expectation of the latent variable conditional on covariates. Hence, we avoid specifying outcome transformations and separate estimand-specific assumptions. We introduce the method; describe identification, estimation, and inference; conduct simulations evaluating its performance; and apply it to assess the impact of Medicaid expansion on CD4 count among people with HIV.\par
\end{abstract}

\section{Introduction}\label{s:intro}

Difference-in-differences (DID) is a popular approach to estimate causal effects with observational data. In its simplest form, DID compares the difference in some mean outcome within a group before and after an intervention to the difference in mean outcome in a group followed over the same time that did not experience the intervention (i.e., a control group).\cite{Wing} The use of DID in this two-group, two-time period setup is well understood and has been used and studied extensively.\cite{Wing,Dimick2014MethodsFE} There have been numerous extensions of the DID approach to accommodate various real-world scenarios, including covariate adjustment, multiple time periods, and multiple intervention arms.\cite{ROTH2023} In this paper, we use the word ``treatment" to represent any intervention or policy for simplicity. \par

The traditional DID approach estimates the average treatment effect among the treated (ATT), which is the effect of the treatment on the mean outcome among those that actually experienced the treatment. Valid estimation of the ATT using traditional DID approaches relies on the parallel trends assumption, which states that the change over time in mean outcome of those in the treatment group, had they been untreated, would be the same as the change over time in mean outcome for those in the control group, possibly conditional on covariates. The ATT and the parallel trends assumption are defined with respect to differences in means. In the presence of skewness or outliers, estimators based on means may suffer from instability and sensitivity to extreme values. It is thus common practice to rely on a transformation, such as the log transformation, with skewed outcome data. However, transforming skewed outcome data in traditional DID approaches leads to other challenges. First, the choice of transformation is sometimes difficult to select and results are often sensitive to that choice. Second, interpretation after transformation may be challenging; one cannot simply back-transform an estimate to obtain the ATT on the original scale. Further, applying a DID approach to transformed outcome data requires the parallel trends assumption to hold on the scale of the mean transformed outcome.\cite{roth_when_2023}\par

In practice, a researcher may be interested in estimating the quantile treatment effect among the treated (QTT) or the probability treatment effect among the treated (PTT). The QTT characterizes the impact of a treatment on a specific quantile of the outcome, while the PTT corresponds to the impact of a treatment on the probability that the outcome is less than or equal to a particular value, both among the treated. These estimands provide additional insights, are applicable to ordinal data and data with detection limits, and are appealing because they avoid complications surrounding estimating an ATT with skewed data or outliers. Estimation of the QTT and PTT within the context of DID has been studied by others.\cite{athey_identification_2006,bonhomme_recovering_2011,Callaway_QTT_2019,CALLAWAY2018} For example, Callaway \textit{et al}.\cite{CALLAWAY2018} developed an approach to estimate the QTT in the setting of two groups and two time periods, making a parallel trends-type assumption on the distributions.\par

Rank-based estimands can also be considered as alternatives to average treatment effects. The Mann-Whitney/Wilcoxon test is commonly used as a non-parametric alternative to the two-sample \textit{t}-test as a measure of stochastic ordering rather than location shift. Many have advocated using the Mann-Whitney parameter, also known as the probabilistic index, as a clinically interpretable measure quantifying the difference between two groups.\cite{acion_probabilistic_2006, thas2012} Specifically, the Mann-Whitney parameter quantifies the probability of a randomly selected treated individual having a higher outcome value than that of a randomly selected control individual. Mann-Whitney-type causal effects have been defined and estimated using rank-based approaches.\cite{fay_causal_2018,zhang_estimating_2019} However, to our knowledge, a Mann-Whitney treatment effect among the treated (MTT) has not yet been considered in DID settings.\par

Given that each estimand can provide specific insights, a researcher may be interested in simultaneously estimating several of the treatment effects mentioned above to get a more comprehensive understanding of the overall impact of a treatment. However, multiple estimands correspond to multiple assumptions, with no guarantee that these assumptions are compatible. For example, use of DID to estimate an ATT involves a parallel trends assumption on the means.\cite{Wing} On the other hand, use of DID to estimate a QTT involves a parallel trends assumption on the quantiles or the entire distribution.\cite{Callaway_QTT_2019} One may also consider using a DID approach to estimate a PTT by dichotomizing the outcome and making a parallel trends assumption on the mean of the dichotomized outcome.\par

In this paper we develop a DID method that permits estimation of the ATT, QTTs, PTTs, and MTT using a distributional regression approach that makes a single unifying parallel trends assumption. Our proposed method involves fitting a semi-parametric cumulative probability model (CPM).\cite{liu_modeling_2017} A CPM is a linear model for a latent outcome variable, which is a transformation of the observed outcome but with the transformation unspecified. The CPM can also be cast as a model on a link function transformation of the cumulative distribution function. This model is particularly suitable for outcome data that are skewed, ordinal, or mixed (e.g., data with detection limits\cite{TianDetection2024}). Our parallel trends assumption is made on the expectation of the outcome on the latent variable scale. The robust, rank-based CPM is compatible with our robust, rank-based estimands. Our proposed method does not require a separate set of assumptions to identify and estimate each causal estimand. In Section \ref{Estimand Definitions}, we define the estimands. 

The rest of the paper is organized as follows. In Section \ref{Identification}, we define assumptions and identify the estimands. In Section \ref{Estimation and Inference}, we describe estimation and inference. We conduct a simulation study in Section \ref{Simulation} to evaluate the performance of our approach. In Section \ref{Application}, we present an application of our approach to estimate the impact of Medicaid expansion on CD4 count at enrollment into care among people living with HIV. Lastly, Section \ref{Discussion} contains a discussion. \par

\section{Estimands}\label{Estimand Definitions}

We consider the scenario of two groups and two time periods. We define the two time periods as the pre-treatment period, $T=0$, and the post-treatment period, $T=1$. We define the two groups as the control group, $D=0$, and the treatment group, $D=1$. Let $ A = 0/1$ denote assignment to no treatment/treatment, respectively. During $T=0$, neither group receives the treatment ($A=0$) and during $T=1$, only those in group $D=1$ receive the treatment ($A=1$). Our observed continuous outcome is denoted by $Y_{dt}$, where the subscripts refer to group $D = d$ and time $T = t$. We allow for the possibility of longitudinal data (i.e., an individual can have an observation during $T=0$ and/or $T=1$) or independent data (i.e., an individual can have an observation at $T=0$ or $T=1$, but not both). Further, we define $Y_{1|dt}$ as the potential outcome for someone in group $d \in \{0, 1\}$ at time $t \in \{0, 1\}$ under the hypothetical scenario that an individual is assigned treatment $A = 1$ and define $Y_{0|dt}$ as the potential outcome under the hypothetical scenario that an individual is assigned control $A = 0$. Additionally, we define $X$ as a covariate or vector of covariates that are unaffected by treatment; $X$ can include interaction terms, splines, etc.\par
We are interested in estimating four marginal treatment effects: the ATT, the QTT, the PTT, and the MTT. The ATT is defined as $\text{ATT} = \text{E}\left[Y_{1|11} - Y_{0|11}\right]$, the mean difference between the treated and untreated potential outcomes among all individuals in the treatment group during the post-treatment period. The ATT can be defined in terms of the cumulative distribution functions (CDF) of the treated and untreated potential outcomes as 
\begin{equation}\label{att}
\text{ATT} = \int ydF_{Y_{1|11}}(y) - \int ydF_{Y_{0|11}}(y),
\end{equation}
\noindent where $F_{Y_{1|11}}(y)$ is the CDF of the treated potential outcome among those in group $D=1$ during time period $T=1$ and $F_{Y_{0|11}}(y)$ is the CDF of the untreated potential outcome among those in group $D=1$ during time period $T=1$. We can also define QTT($p$) for any probability $p \in (0,1)$ as
\begin{equation}\label{qtt}
\text{QTT}(p) = F^{-1}_{Y_{1|11}}(p) - F^{-1}_{Y_{0|11}}(p),
\end{equation}
\noindent where $F^{-1}_{Y_{1|11}}(p)$ and $F^{-1}_{Y_{0|11}}(p)$ represent the $p^{\text{th}}$ quantiles of $F_{Y_{1|11}}(y)$ and $F_{Y_{0|11}}(y)$, respectively. We define PTT($y$) for any possible value of $Y = y$ as 
\begin{equation}\label{ptt}
\text{PTT}(y) = F_{Y_{1|11}}(y) - F_{Y_{0|11}}(y).
\end{equation}
\noindent  Note that the definitions above for ATT, QTT, and PTT involve subtraction; other contrasts could be used (e.g., ratio). Lastly, we define the MTT as
\begin{equation}\label{mtt}
\text{MTT} = \iint h(u,v)dF_{Y_{1|11}}(u)dF_{Y_{0|11}}(v).
\end{equation}
\noindent where $ h(u,v) = I\{u >v\} + 0.5I\{u=v\}$ with $I\{\cdot\}$ denoting the indicator function.\cite{zhang_estimating_2019} Therefore, the $MTT = \text{E}\left[h\left(Y_{1|11i}, Y_{0|11j}\right)\right]$ 
where subscripts $i \ne j$ denote two independent and randomly selected subjects from the same target population with $D=1$ and $T=1$; $Y_{1|11i}$ denotes the potential outcome for subject $i$ if assigned treatment and $Y_{0|11j}$ denotes the potential outcome for subject $j$ if assigned control. The MTT for a continuous outcome is interpreted as the probability that a randomly drawn individual in the treatment group would have a higher outcome as compared to another randomly drawn individual in the treatment group had they not been treated.\par
Note that the  ATT, QTT, PTT, and MTT are marginal treatment effects that are defined on the potential outcome scale. The potential outcomes $Y_{1|11}$ and $Y_{0|11}$ cannot both be directly observed for an individual during any one period. However, information about the joint distribution of $Y_{1|11}$ and $Y_{0|11}$ is not needed for our estimands. For example, the QTT, as defined in (\ref{qtt}), compares the $p^{\text{th}}$ quantile of $Y_{1|11}$ versus the $p^{\text{th}}$ quantile of $Y_{0|11}$, not the $p^{\text{th}}$ quantile of $Y_{1|11} - Y_{0|11}$; the ATT and PTT are similarly defined with marginal distributions. Similarly, the MTT is defined such that $Y_{1|11}$ and $Y_{0|11}$ are required to be drawn from separate (and independent) subjects. Therefore, the joint distribution of $Y_{1|11i}$ and $Y_{0|11j}$ factors into two marginal distributions, as shown in (\ref{mtt}). Because the four estimands only rely on the two marginal distributions of the potential outcomes rather than their joint distribution, identification of the estimands requires fewer assumptions than otherwise. Issues surrounding identifiability of joint versus marginal distributions have been discussed by others.\cite{fay_causal_2018}

\section{Identification}\label{Identification}
\subsection{Assumptions}\label{assumptions}
We now introduce assumptions to identify the estimands defined in the previous section.

\begin{assumption}[\textbf{Consistency}]\label{assump:cons}
	$Y_{a\mid dt}=Y_{dt}$ if $A=a$.
\end{assumption}

\begin{assumption}[\textbf{No Interference}]\label{assump:no_int}
	The potential outcomes $Y_{0|dt}$ and $Y_{1|dt}$ for an individual are not affected by the treatment assignment and covariates $(A,D,T,X)$, of other individuals in the study.
\end{assumption}

\begin{assumption}[\textbf{Conditional Latent Parallel Trends}]\label{assump:cond_pta} 
$G\{F_{Y_{0|11}}(y|X)\} - G\{F_{Y_{0|10}}(y|X)\} = G\{F_{Y_{0|01}}(y|X)\} - G\{F_{Y_{0|00}}(y|X)\}$
\end{assumption}

\noindent where $G(\cdot)$ is a link function and $F_{Y_{a|dt}}(y|X) = P(Y_{a|dt} \leq y|X)$.\par
Assumption \ref{assump:cons} states that the potential outcome of an individual under their observed treatment status is the observed outcome for that individual. \ref{assump:cons} can be violated if there are multiple manners in which the treatment was implemented.\cite{cole_frangakis_2009} Assumption \ref{assump:cons} also implicitly imposes a non-anticipation assumption, $Y_{d0} = Y_{0|d0}$, that the pre-treatment outcome is not affected by upcoming treatment. This assumption could be violated if, for example, someone altered their behavior during the pre-treatment period based on their knowledge/beliefs about the treatment that would occur during the post-treatment time period. Assumption \ref{assump:no_int}, sometimes referred to as no spillover effects, allows us to write the potential outcomes using the treatment status of an individual, rather than for all individuals in the study. \ref{assump:no_int} is plausible in settings where there is no interaction between study subjects. Assumptions \ref{assump:cons} and \ref{assump:no_int} constitute the popular SUTVA (stable unit treatment values assumption).\cite{rubin2005} Assumption \ref{assump:cond_pta} says that after a link function transformation, the difference between the distributions of potential outcomes for $A=0$ in the post- / pre- treatment period is the same in the treatment and control groups; this assumption will be discussed in more detail after we present our model. \par
Skewed continuous response variables are common in practice, and they are often transformed prior to fitting a linear model. For example, skewed data may be log-transformed and then a regression model is fit to the log-transformed outcome. However, the transformation may be difficult to choose and results can be sensitive to the choice of transformation. Semi-parametric linear transformation models assume that after a monotonic transformation the transformed outcome follows a linear model – but they avoid specifying the functional form of the transformation.\cite{zeng_maximum_2007} We use these models in our DID analysis framework. We first express these models in a simple manner, but then describe how they can be generalized. Specifically, we assume

\begin{assumption}[\textbf{Semi-parametric Linear Transformation Model}]\label{assump:ltm}
$Y_{dt} = H(Y_{dt}^*)$, and $Y_{dt}^* = \beta_1d + \beta_2t + \beta_3dt + \beta_4^TX  + \varepsilon$
\end{assumption}

\noindent where $H(\cdot)$ is monotonic increasing but otherwise unspecified and $\varepsilon \sim F_\varepsilon$ where $F_\varepsilon$ is a known distribution.\par
Our assumed model in Assumption \ref{assump:ltm} is semi-parametric because the functional form of
$H(\cdot)$ is not specified and is non-parametrically estimated, while the effect of $D$, $T$, and $X$ are modeled parametrically. The parameter $\beta_3$ can be thought of as the added effect of treatment on the scale of the latent variable, $Y_{dt}^*$. Under \ref{assump:ltm}, the distribution of the outcome conditional on $X$, $T$, and $D$ can be expressed as
\begin{align*}
F(y|D,T,X) &= P(Y_{dt} \leq y|D,T,X)\\
& = P\left(H(\beta_1d + \beta_2t + \beta_3dt + \beta_4^TX + \varepsilon) \leq y\right)\\
&= P(\beta_1d + \beta_2t + \beta_3dt + \beta_4^TX + \varepsilon \leq H^{-1}(y)| X)\\
& = F_{\varepsilon}\left(H^{-1}(y)-\beta_1d - \beta_2t - \beta_3dt - \beta_4^TX\right).
\end{align*}
After rearrangement, we have our cumulative probability model (CPM),
\begin{equation}\label{cpm}
G\left[P\left(Y_{dt}\leq y|X\right)\right] =   H^{-1}(y) -\beta_1d - \beta_2t - \beta_3dt - \beta_4^TX,
\end{equation}
\noindent where $G(\cdot) = F^{-1}_{\varepsilon}(\cdot)$ is a link function. CPMs are very flexible and robust because they non-parametrically estimate $H^{-1}(y)$ in a rank-based manner, as will be detailed in Section \ref{Estimation and Inference}. These models have been studied with ordinal outcomes,\cite{McCullagh} time-to-event outcomes,\cite{zeng_maximum_2007} continuous outcomes, \cite{liu_modeling_2017} and mixtures of ordinal and continuous outcomes.\cite{TianDetection2024}\par
For simplicity and clarity in presentation, we have chosen to include $D$, $T$, and $X$ as in (\ref{cpm}). However, as with other regression models, CPMs could include functions of $X$ (e.g., polynomials, splines, tree partitions, products) and interaction terms between functions of $X$ and $D$, $T$. More generally, we could write the model as 
\begin{equation}\label{gen}
Y^*_{dt} = \gamma_1^T \eta_1(X)d + \gamma_2^T \eta_2(X)t + \gamma_3^T \eta_3(X)dt + \gamma_4^T \eta_4(X) + \varepsilon,
\end{equation}
where $\eta_1(\cdot),\eta_2(\cdot),\eta_3(\cdot)$ and $\eta_4(\cdot)$ are pre-specified functions of $X$ and $\gamma_1, \gamma_2,\gamma_3$ and $\gamma_4$ are parameters of compatible dimensions. All the identification, estimation, and inference results that will be presented hereafter can be easily written in an analogous manner under the more general model (\ref{gen}). Note also that CPMs are a special type of distributional regression model,\cite{chernozhukov_inference_2013} which have become popular in the economics literature. Unlike general distribution regression models, however, we do not permit $\gamma_1, \gamma_2,\gamma_3$ or $\gamma_4$ (nor $\beta_1,\beta_2,\beta_3$, or $\beta_4$) to be functions of the outcome, $y$. Thus, our model is less flexible than these distribution regression models, but more parsimonious. As we will illustrate by simulations in Section \ref{Simulation}, parsimony leads to improvements in efficiency. An additional advantage of the $\gamma$'s (and $\beta$'s) not depending on $y$ is that the CDFs from our model are monotonic and thus valid CDFs.

Analogous to the potential outcomes on the original scale, $Y_{1|dt}^* = H^{-1}(Y_{1|dt})$ and $Y_{0|dt}^*= H^{-1}(Y_{0|dt})$ are the potential outcomes on the latent variable scale. Note that assumption \ref{assump:cons} also holds for the latent variables. Our CPM allows us to cast assumption \ref{assump:cond_pta} in a more interpretable manner:

\begingroup
\renewcommand{\theassumption}{A3$'$}
\begin{assumption}[\textbf{Conditional Latent Variable Parallel Trends}]
	\label{assump:cond_pta_prime}
	$\text{E}[Y^{*}_{0\mid11}-Y^{*}_{0\mid10}\mid X=x]
	=\text{E}[Y^{*}_{0\mid01}-Y^{*}_{0\mid00}\mid X=x]$
\end{assumption}
\endgroup

\noindent Assumption \ref{assump:cond_pta_prime} states that for individuals with the same observed covariate values, the change over time in mean latent outcome of those in the treatment group, had they not been treated, would be the same as the change over time in the control group. This is a variation of the parallel trends assumption required in DID approaches, the difference being we are making the assumption on the latent variable scale, not the original scale. The parallel trends assumption on either scale is not directly testable in two-group, two-time period DID approaches. However, if there are multiple measures over time in the pre-treatment period, then the plausibility of \ref{assump:cond_pta}/\ref{assump:cond_pta_prime} can be investigated. Specifically, one could fit a CPM in the pre-treatment period; evidence of a time by group interaction in the pre-treatment period would suggest that \ref{assump:cond_pta_prime} is unlikely to hold. In any case, it is important to think conceptually about why \ref{assump:cond_pta_prime} may or may not be valid. For example, are there other differences (e.g., policy changes or secular trends) between the treatment and control groups that may be differently affecting outcome trends over time?

\subsection{Identification of Estimands}\label{Identification of Estimands}
The  ATT, QTT, PTT, and MTT are all defined using the marginal distributions of the potential outcomes. Therefore, to identify these estimands, we first identify $F_{Y_{1|11}}(y)$ and $F_{Y_{0|11}}(y)$ with observable data using the assumptions from the previous section.\\

\begin{identres}\label{idres:1}
Under Assumptions \ref{assump:cons}-\ref{assump:no_int}, it is straightforward to show that the marginal distribution of $Y_{1|11}$ is $F_{Y_{1|11}}(y) = \int P\left(Y_{11} \leq y |D=1\right)dF_{X}(x|D=1,T=1).$ Under \ref{assump:ltm}, this can be written as $\int F_{\varepsilon}\left(H^{-1}(y) - \beta_1 - \beta_2 - \beta_3 - \beta_4^Tx \right)dF_{X}(x|D=1,T=1).$\\
\end{identres}

\begin{identres}\label{idres:2}
Under Assumptions \ref{assump:cons}, \ref{assump:no_int}, \ref{assump:cond_pta}, and \ref{assump:ltm}, the marginal distribution of $Y_{0|11}$ is $F_{Y_{0|11}}(y) =\int F_{\varepsilon}\left(H^{-1}(y) - \beta_1 - \beta_2 - \beta_4^Tx\right)dF(x|D=1,T=1).$\\
\end{identres}

\noindent The proof of Identification Result \ref{idres:2} can be found in the Supplemental Material.\par

Under Assumptions \ref{assump:cons}-\ref{assump:ltm} and Identification Results \ref{idres:1} and \ref{idres:2}, expressions (\ref{att}) - (\ref{mtt}) are identifiable and thus ATT, QTT, PTT, and MTT can be written in terms of observable data.
\section{Estimation and Inference}\label{Estimation and Inference}
\subsection{Estimation}\label{Estimation}
We first describe the estimation of $\beta_1$, $\beta_2$, $\beta_3$, $\beta_4$, and $H^{-1}(y)$ from model (\ref{cpm}), the components that comprise the marginal distributions of the potential outcomes. Estimation proceeds in a similar manner for the more general model (\ref{gen}). For simplicity in presentation, we assume there are no ties in the observed outcomes, denoted as $y_i$ $(i = 1, \ldots, n)$; extensions to settings with ties are straightforward.\cite{TianDetection2024} Without loss of generality, we further assume $y_1 < y_2 < \ldots < y_n$ and $H^{-1}(y)$ is strictly increasing. Let $\alpha_i = H^{-1}(y_i)$; note that $\alpha_1 < \alpha_2 < \ldots < \alpha_{n}$. Following Liu \textit{et al.},\cite{liu_modeling_2017} the non-parametric likelihood for the CPM for independent and identically distributed realizations of $(Y_i, D_i, T_i, X_i)$ is
\begin{align*}
L^*(\beta, \alpha) &= \prod_{i = 1}^{n} \left[G^{-1}\left(\alpha_i - \beta^TW_i\right) - G^{-1}\left(\alpha_{i-1} - \beta^TW_{i}\right)\right],
\end{align*}
\noindent where $\beta^TW_{i} = \beta_1D_i + \beta_2T_i + \beta_3D_iT_i + \beta_4^TX_i$ and $\alpha_0$ ($<\alpha_1$) is an auxiliary parameter. The likelihood is maximized when $\widehat{\alpha}_{0} = - \infty$ and $\widehat{\alpha}_{n} = \infty$, so these parameters are fixed and the likelihood can be written as
\begin{align}\label{npl}
L^*(\beta, \alpha) &= \left[G^{-1}\left(\alpha_1 - \beta^TW_1\right)\right]\notag\\
&\ \ \ \ \times \prod_{i = 2}^{n-1} \left[G^{-1}\left(\alpha_i - \beta^TW_i\right) - G^{-1}\left(\alpha_{i-1} - \beta^TW_{i}\right)\right]\\
&\ \ \ \ \times \left[1 - G^{-1}\left(\alpha_{n-1} - \beta^TW_n\right)\right].\notag
\end{align}
Non-parametric maximum likelihood estimators (NPMLE) for $\beta_1$, $\beta_2$, $\beta_3$, $\beta_4$, and ($\alpha_1, \ldots, \alpha_{n-1}$) are obtained by maximizing the likelihood in equation (\ref{npl}). The intercept function $H^{-1}(y)$ is thus estimated as a step function. Specifically, for any $y \in [y_{min}, y_{max}]$,  $\widehat{H}^{-1}(y) = \widehat{\alpha}_{i_y}$ where $i_y = max\{i: y_i \leq y\}$; when $y < y_{min}$, $\widehat{H}^{-1}(y) = -\infty$ and when $y \geq y_{max}$, $\widehat{H}^{-1}(y) = \infty$. Obtaining the NPMLEs can be done by treating the continuous outcome as a discrete variable (i.e., each outcome value is its own category) and fitting an ordinal CPM (also referred to as a cumulative link model \cite{agresti_analysis_2010}). This can be done in a computationally efficient manner using the \texttt{orm} function in the \texttt{rms} package \cite{rms} in R.\par
Although (\ref{npl}) is defined for independent outcomes, our proposed method allows outcomes to be clustered at the individual level (i.e., an individual can have an observation during $T=0$ and/or $T=1$). Tian \textit{et al.}\cite{Tian2023} extend the estimation of $\beta$ coefficients and ($\alpha_1, \ldots, \alpha_{n-1}$) to the setting of clustered continuous outcomes by fitting a CPM with generalized estimating equation (GEE) methods for ordinal data. Parameter estimates with clustered continuous outcome data with an independence working correlation structure are the same as those from Liu \textit{et al.}\cite{liu_modeling_2017} for independent data. The variance can be estimated with a sandwich variance estimator. Therefore, with an independence working correlation structure, we can obtain point estimates for $\beta_1$, $\beta_2$, $\beta_3$, $\beta_4$, and ($\alpha_1, \ldots, \alpha_{n-1}$) by following the above procedure for independent continuous outcome data. In this paper, we do not detail the estimation procedure of our treatment effects with other working correlation structures, although estimation procedures exist in such settings.\cite{Tian2023}\par

After we estimate $\beta_1$, $\beta_2$, $\beta_3$, $\beta_4$, and $(\alpha_1, \ldots, \alpha_{n-1})$ by maximizing (\ref{npl}), we can estimate $F_{Y_{1|11}}(y)$ and $F_{Y_{0|11}}(y)$ using Identification Results \ref{idres:1} and \ref{idres:2}, replacing the right hand sides of these equations with plug-in estimators. Then estimation of the ATT, QTT, PTT, and MTT follow. Specifically, $F_{Y_{1|11}}(y)$ can be estimated as the following:
\begin{align*}
\widehat F_{Y_{1|11}}(y) &= \sum_{x \in \mathcal{X}}F_{\varepsilon}\left(\widehat{H}^{-1}(y) - \widehat \beta_1 - \widehat \beta_2 - \widehat \beta_3 - \widehat\beta_4^Tx \right)\\
& \ \ \ \ \ \times \widehat P(X = x| D=1, T=1),
\end{align*}
\noindent where $\mathcal{X}$ is the set of all observed covariates in the data, $\widehat P(X = x| D=1, T=1) = \frac{1}{n_{11}}\sum_{i = 1}^{n} I\{X_i =x\}I\{D_i =1, T_i = 1\}$, $n$ is the total sample size, and $n_{11} = \sum_{i =1}^{n} I\{D_i = 1, T_i = 1\}$ is the number of individuals in group $D=1$ during time period $T=1$. Similarly, $F_{Y_{0|11}}(y)$ can be estimated as the following:
\begin{align*}
\widehat F_{Y_{0|11}}(y) &= \sum_{x\in \mathcal{X}}F_{\varepsilon}\left(\widehat{H}^{-1}(y) - \widehat \beta_1 - \widehat \beta_2 -  \widehat\beta_4^Tx \right) \widehat P(X = x| D=1, T=1).
\end{align*}
While the distributions of the potential outcomes are now estimated, estimation of the  ATT requires further consideration. We first note that $\text{E}[Y_{a|11}] = \int y dF_{Y_{a|11}}(y)$ can be approximated by $\sum_{i = 1}^{n}y_i\left[\widehat F_{Y_{a|11}}\left(y_i\right) -\widehat F_{Y_{a|11}}\left(y_{i-1}\right)\right]$ for $a = 0,1$, where $y_0 = - \infty$ and $\widehat F_{Y_{a|11}}(y_0) =0$. The ATT can then be estimated as 
\begin{align*}
\widehat{\text{ATT}} &= \sum_{i = 1}^{n}y_i\left[\widehat F_{Y_{1|11}}\left(y_i\right) - \widehat F_{Y_{1|11}}\left(y_{i-1}\right)\right]\\
& \ \ \ \ - \sum_{i = 1}^{n}y_i\left[\widehat F_{Y_{0|11}}\left(y_i\right) - \widehat F_{Y_{0|11}}\left(y_{i-1}\right)\right].
\end{align*}
The estimate of the quantile treatment effect for the $p^{\text{th}}$ quantile is
\begin{align*}
\widehat{\text{QTT}}(p) = \widehat F^{-1}_{Y_{1|11}}(p) - \widehat F^{-1}_{Y_{0|11}}(p).
\end{align*}
\noindent  For $a=0,1$, $\widehat F^{-1}_{Y_{a|11}}(p)$ can be estimated through linear interpolation following the procedure described in Tian \textit{et al}.\cite{TianDetection2024}\par
An estimate for PTT follows directly from the estimation of the distributions of the potential outcomes.
\begin{align*}
\widehat{\text{PTT}}(y)&= \widehat F_{Y_{1|11}}(y) - \widehat F_{Y_{0|11}}(y).
\end{align*}
\noindent Similar to the estimation of the ATT, the estimation of the MTT uses the fact that $d {{F}}_{Y_{a|11}}(y_i)$ can be approximated as $\widehat{F}_{Y_{a|11}}(y_i) - \widehat{F}_{Y_{a|11}}(y_{i-1})$ for $a = 0,1$ so that
\begin{align*}
\widehat{\text{MTT}} &= \underset{j\ne i}{\sum_{i = 1}^n \sum_{j=1}^n}h(y_{i}, y_{j}) \left[ \widehat F_{Y_{1|11}}(y_i) - \widehat F_{Y_{1|11}}(y_{i-1}) \right]\\
&\ \ \ \ \times \left[\widehat F_{Y_{0|11}}(y_j) - \widehat F_{Y_{0|11}}(y_{j-1}) \right].
\end{align*}
\noindent The double summation indicates we are considering every pairwise comparison of outcomes for independent individuals in the data.

\subsection{Inference}\label{Inference}
The asymptotic properties of the estimated transformation function $H^{-1}(y)$ and the estimated $\beta$ coefficients for semi-parametric transformation models have been established in the setting of censored outcome data, which relies on the boundedness of the transformation function.\cite{zeng_maximum_2007} More recently, Li \textit{et al.}\cite{li_asymptotic_2023} showed that under mild conditions, CPMs for continuous outcome data result in consistent and asymptotically normal estimates of $H^{-1}(y)$ and $\beta$. Specifically, they avoid complications surrounding the potentially unbounded range of the transformation by modifying the continuous outcome data such that values below a pre-specified lower bound ($L$) or above a pre-specified upper bound ($U$) are censored, resulting in two ordinal categories, representing the lowest and highest ranked values, respectively. Li \textit{et al.}\cite{li_asymptotic_2023} performed extensive simulations comparing estimators with and without modifying the data and illustrated that, provided that the fraction of the data being censored was small (e.g., $<$5-10\%), the estimated $\beta$ and $H^{-1}(y)$ for $y$ in ($L$,$U$) were almost identical, and the estimated conditional mean without censoring appeared to be consistent. \par

We now consider large-sample properties of our estimators under the same data modification and conditions as Li \textit{et al}.\cite{li_asymptotic_2023} First note that $\widehat P(X = x| D=1, T=1) =  \frac{1}{n_{11}}\sum_{i = 1}^{n} I\{X_i =x\}I\{D_i =1, T_i = 1\}$ is a sample mean, thus we have that $\widehat P(X = x| D=1, T=1) \rightarrow P(X = x| D=1, T=1)$. We also have that $\widehat{H}^{-1}(y)$ and $\widehat \beta$ are consistent and asymptotically normal for $y$ $\in$ ($L$,$U$). Thus, $\widehat F_{Y_{1|11}}(y)$ is also asymptotically normal and consistent. By the same logic, $\widehat F_{Y_{0|11}}(y)$ is also asymptotically normal and consistent for $y$ $\in$ ($L$,$U$). Thus $\widehat{\text{PTT}}$$(y)$ is asymptotically normal and consistent for $y$ $\in$ ($L$,$U$). $\widehat{\text{QTT}}$($p$) is also asymptotically normal and consistent for $p$ sufficiently far from 0 and 1, such that the corresponding quantile falls within ($L$,$U$). In contrast, the consistency of estimators of the ATT and MTT require consistent estimation of the entire distribution, not just within ($L$,$U$), which is not guaranteed by the results of Li \textit{et al.}\cite{li_asymptotic_2023}\par 
We note that our estimation procedure has not required these bounds on the outcome. As with Li \textit{et al.}\cite{li_asymptotic_2023}, our simulations (see next section) suggest that our estimators of the ATT, QTT, PTT, and MTT are consistent and asymptotically normal when the data are not censored.\par 
We compute 95\% confidence intervals using standard percentile bootstrap procedures. For data with more than one observation per subject, a cluster bootstrap is used. Other types of bootstrap CIs could be constructed.\cite{DiCiccio_Efron_1996}

\section{Simulation Study}\label{Simulation}
We performed several sets of simulations evaluating the performance of our estimators under different scenarios. We first describe the general data generation mechanism, and then we describe simulation results under specific settings.
\subsection{Data Generation}
For all simulation scenarios, we generated two-group, two-period settings with approximately half of subjects having observations in both time periods. We first generated whether a subject would have one or two measurements from a Bernoulli(0.5) distribution. For those subjects with one observation, we then generate data from $D \sim$ Bernoulli(0.5), $T \sim$ Bernoulli(0.5), and $\varepsilon_{0T}$ and $\varepsilon_{1T}$ from independent $\text{Normal}(0,1)$. For those with two observations, we generate $D \sim$ Bernoulli(0.5) and $\left(\varepsilon_{00}, \varepsilon_{01}\right)$ and $\left(\varepsilon_{10}, \varepsilon_{11}\right)$ from $\text{Normal}\bigl(\bigl(\begin{smallmatrix}0\\ 0\end{smallmatrix}\bigr),\left(\begin{smallmatrix}1 & 0.5\\ 0.5 & 1\end{smallmatrix}\right)\bigr)$. We next generated two covariates, $X = \left(X_1,X_2\right)$, where $X_1\sim \text{Bernoulli}(0.5)$ and $X_2\sim \text{Normal}(0,1)$. 

We next generated potential outcomes on the latent variable scale, $Y_{a|dt}^*$.  We generated potential outcomes without treatment (i.e., $Y_{0|00}^*$, $Y_{0|01}^*$, $Y_{0|10}^*$, and $Y_{0|11}^*$) as $Y_{0|dt}^* = \beta_1 d + \beta_2 t + \beta_4^TX + \varepsilon_{dt}$. The potential outcomes under treatment were generated as $Y_{1|11}^*=Y_{0|11}^*+\beta_3$. We set $\beta_1 = 1$, $\beta_2 = 0.5$, $\beta_3 = 0.5$, $\beta_4 = (0.25,0.5)$. We next generated the observed data, $Y_{dt}$, by transforming the relevant potential outcomes. Specifically, $Y_{00}=H(Y^*_{0|00})$,  $Y_{01}=H(Y^*_{0|01})$, $Y_{10}=H(Y^*_{0|10})$, and $Y_{11}=H(Y^*_{1|11})$.  This set-up satisfies assumptions \ref{assump:cons}-\ref{assump:cond_pta}. In our primary simulation setting, we set $H(\cdot) = \exp(\cdot)$ so that realized outcomes on the original scale are $Y_{dt} = \exp(Y^{*}_{dt})$. We consider this transformation to evaluate the performance of our proposed method in a common situation where the outcome is right skewed. In a few other simulation settings, we also generated $Y_{dt}$ using the identity transformation, $H(y)=y$. In all settings, our estimator did not know the transformation; $H(\cdot)$ was estimated non-parametrically as part of the estimation as outlined in Section \ref{Estimation and Inference}.

We conducted simulations for sample sizes of $n$ = 200, 300, 1000, 1500, and 2000 where $n$ is the number of subjects. We generated 1000 replicates for each sample size and for variance estimation we used 200 bootstrap samples per replicate. Two hundred bootstrap samples were chosen to reduce the computational burden.

To evaluate the performance of our estimators, we first generated a pseudo population of 10$^7$ observations using the data generating process above and then empirically estimated the true values for our population in each scenario. The true values for the  ATT and PTT were calculated as $\text{ATT} = \widehat{\text{E}}[H(Y^*_{1|11})  - H(Y^{*}_{0|11}) ]$ and $\text{PTT}(y) = \widehat{\text{E}}\left[I\{H(Y^*_{1|11})\leq y\} \right] - \widehat{\text{E}}\left[I\{H(Y^*_{0|11})\leq y\} \right]$ for any value of $y$ where $\widehat{\text{E}}[\cdot]$ is the empirical expectation. To calculate a true QTT($p$) value for any value of $p$, we subtracted the $p^{\text{th}}$ quantile of $H(Y^{*}_{0|11})$  from the $p^{\text{th}}$ quantile of $H(Y^*_{1|11})$. Lastly, to calculate the true value of the MTT, we randomly sampled 10$^7$ sets of two independent individuals, $i$ and $j$ such that $i \ne j$, to calculate $\widehat{\text{E}}\left[h(H(Y^*_{1|11i}), H(Y^*_{0|11j}))\right]$. We focus on the following estimands, which are approximately $\text{ATT} = 6.2$, $\text{QTT}(0.25) = 1.5$, $\text{QTT}(0.50) = 3.3$, $\text{QTT}(0.75) = 7.0$, $\text{PTT}(1) = -0.045$, $\text{PTT}(3) = -0.139$, $\text{PTT}(6) = -0.175$, and $\text{MTT} = 0.623$.

\subsection{Simulation Results}

\subsubsection{Performance with proper model specification}
\begin{figure}[!ht]
	\centering
	\includegraphics[width=5cm]{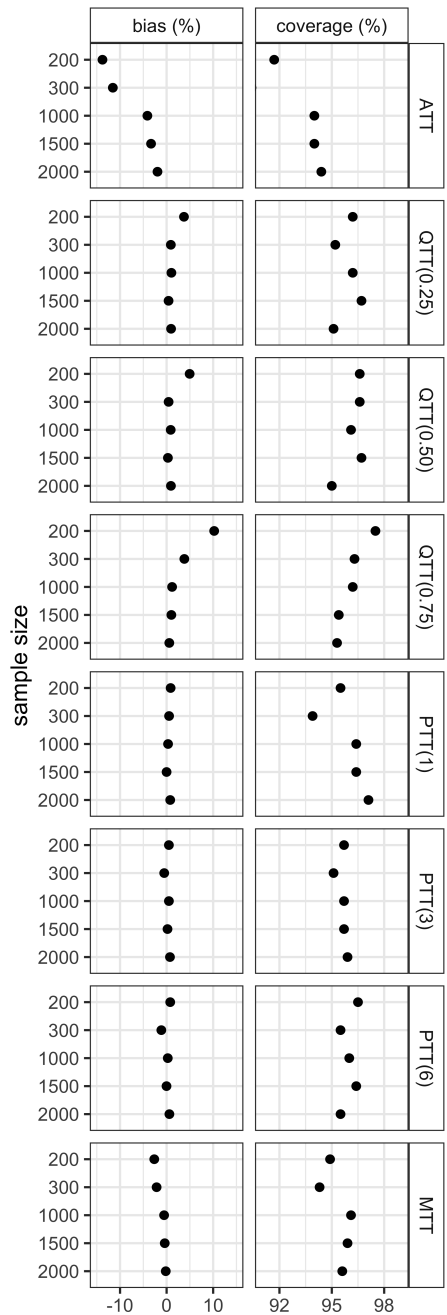}
		\caption{Percent bias and coverage of ATT, QTT$(0.25)$, QTT$(0.50)$, QTT$(0.75)$, PTT$(1)$, PTT$(3)$, PTT$(6)$, and MTT estimators from simulation of 1000 replications with 200 bootstrap iterations (as described in Section \ref{Inference}) for the scenario in which $Y = \exp(D + 0.5T + 0.5DT + 0.25X_1 + 0.5X_2 + \varepsilon_T)$, and $\varepsilon_T \sim \text{Normal}(0,1)$, with link function specified as probit.$^*$ \\\footnotesize{*ATT (average treatment effect among the treated), QTT (quantile treatment effect among the treated), PTT (probability treatment effect among the treated), MTT (Mann-Whitney treatment effect among the treated).}}
		\label{fig:sim_results_main}
\end{figure}

We summarize the performance of $\widehat{\text{ATT}}$, $\widehat{\text{QTT}}(0.25)$, $\widehat{\text{QTT}}(0.50)$, $\widehat{\text{QTT}}(0.75)$, $\widehat {\text{PTT}}(1)$, $\widehat{\text{PTT}}(3)$, $\widehat{\text{PTT}}(6)$, and $\widehat{\text{MTT}}$ under correctly specified models (e.g., probit link function) for various sample sizes in Figure \ref{fig:sim_results_main}. Our PTT estimators performed well for all sample sizes considered, with all percent biases $\leq$ 2\%, even with $n$ = 200. While QTT estimators had slightly larger percent biases (up to 10.2\% for QTT(0.75) for $n$ = 200), percent bias for all QTT estimators were $\leq$1.5\% for $n$ $\geq$ 1000. Further, though the percent bias for the ATT estimator was $-$13.8\% for $n$ = 200, percent bias also moved towards zero as sample size increased and was below 2.5\% by sample size $n$ = 1500. The MTT estimator performed relatively well for all sample sizes with percent bias below 1\% by $n$ = 1000. Coverage was close to 95\% for all sample sizes and estimands, except the ATT with $n$ = 200, 300 and the $\text{QTT}(0.75)$ with $n$ = 200. The bias and sub-optimal coverage for the ATT and QTT(0.75) at small sample sizes were as expected, given highly skewed nature of the generated outcome.\par

\subsubsection{Comparison with alternative estimators of ATT}
We next compared our estimator of ATT with two alternative estimators of ATT for data with symmetric residuals. Specifically, we considered three simulation scenarios, all with $Y=H(Y^*) = Y^*$:  I) $\varepsilon \sim $ Normal(0,1) with covariates, II) $\varepsilon \sim $ Normal(0,1) without covariates, and III) $\varepsilon \sim $ Logistic(0,1) with covariates. For each scenario, we compared the performance of our ATT estimator, estimated from properly specified models (e.g., probit link for scenarios I and II, logit link for scenario III; covariates included in scenarios I and III, no covariates in scenario II), against the performance of two alternative ATT estimators, $\widehat{\text{ATT}}'$ and $\widehat{\text{ATT}}''$. $\widehat{\text{ATT}}'$ is the traditional approach used in DID analyses. $\widehat{\text{ATT}}'$ was calculated by fitting a linear model for the outcome on $D$, $T$, an interaction between $D$ and $T$, and covariates (scenarios I and III, no covariates for scenario II); $\widehat{\text{ATT}}'$ is the estimated coefficient for the interaction term. $\widehat{\text{ATT}}''$ is one of the Callaway-Sant'Anna DID estimators.\cite{Callaway} It models the outcome in the control group as a function of time and covariates (scenarios I and III, no covariates in scenario II), uses that model to predict the outcome difference over time in the treatment group in the absence of treatment, and then subtracts the resulting mean difference from the mean difference in the treatment group.\cite{Callaway}  We estimated $\widehat{\text{ATT}}''$ with the \texttt{att\_gt} function in the \texttt{did} package \cite{did_package} in R. 

Whether or not covariates are included in the model, $\widehat{\text{ATT}}'$ implicitly assumes unconditional parallel trends and homogeneous treatment effects.\cite{ROTH2023} In contrast, our estimator and $\widehat{\text{ATT}}''$ make unconditional parallel trends assumptions in scenarios II, but conditional parallel trends assumptions in scenarios I and III. The parallel trends assumptions slightly differ between our estimator and that of $\widehat{\text{ATT}}''$: ours is parallel trends on the means of the latent transformed responses whereas $\widehat{\text{ATT}}''$ is on the means of the original responses. Because the data were generated with an identity transformation function, in theory these are the same; but our estimator does not know that, and it non-parametrically estimates the identity transformation.

In short, except at small sample sizes ($n=200$), our estimator performed similarly to the alternative estimators (Supplemental Figure A1). With $n=200$ under scenarios I and II, our estimators had higher bias; with larger samples sizes, bias was low and comparable with the other estimators. Coverage and relative mean squared error were not markedly different between estimators. These results suggest that with sufficient sample sizes, there are minimal drawbacks in terms of bias, variance, and coverage for the CPM-based ATT estimator, which non-parametrically estimates a transformation when no transformation is needed.

\subsubsection{Comparison with estimators based on distribution regression}
We next compared our CPM-based estimators of the ATT, QTT($p$), and MTT with similar estimators, except plugging-in outcome regression estimates based on distribution regression  (DR) instead of the CPM.\cite{chernozhukov_inference_2013} We considered both $Y = H(Y^*)=\exp(Y^*)$ and $Y = H(Y^*) = Y^*$; for both, $\varepsilon \sim$ Normal(0,1) and other aspects of the data generation were as described in Section 5.1. Both the CPM and the DR estimators correctly assumed a probit link function, and both estimators non-parametrically estimated the transformation $H(\cdot)$. The DR estimators first fit $n-1$ probit regression models, a separate one for every possible  order-maintaining dichotomization of the outcome. Thus, the intercept and $\beta$ coefficients from the DR are functions of the outcome $y$ (i.e., $(n-1)\times3$ parameters); in contrast, for CPMs, only the intercept is a function of the outcome (i.e., $n-1+2$ parameters). The DR-based estimators followed the estimation procedures for the ATT, QTT, PTT, and MTT as described in Section 4.1 except plugging-in estimates from the fitted DR. The DR-based estimators also assume conditional parallel trends (Assumption \ref{assump:cond_pta}). When $Y = \exp(Y^*)$ and $Y = Y^*$, CPM resulted in smaller variances compared to DR for all estimators and sample sizes considered ($n$ = 200, 300, 1000, 1500, 2000, 3000, 5000, and 10,000). Percent bias was also lower for all estimators in all scenarios and sample sizes considered with the CPM-based estimators compared to the DR-based estimators, except for the ATT with $Y = \exp(Y^*)$ and $n$ = 200. The generally better performance of the CPM-based estimators over the DR-based estimators is not surprising, as the extreme flexibility of the DR approach requires estimating many additional parameters which leads to lower efficiency and requires larger sample sizes to reduce bias. Further details are provided in Figures A2 and A3 in the Supplemental Material.\par

\subsection{Performance under moderate link function misspecification}
We also evaluated the performance of our estimators when the link function was misspecified as logit, rather than probit. In both transformation scenarios ($Y=\exp(Y^*)$ and $Y=Y^*)$, our estimators generally performed well for $n \geq 300$, with absolute percent bias $\leq 5\%$ and near-nominal coverage. The notable exception was for estimating the ATT when $Y = \exp(Y^*)$, where absolute percent bias was between 10 to 20\% and coverage $<90\%$. These results generally suggest that our method is fairly robust to moderate link function misspecification. Further details and simulation results are described in Figure A4 in the Supplemental Material.\par

\subsection{Other settings}
Our data generating process set $\varepsilon_{0T}$ and $\varepsilon_{1T}$ as independent, which is typically unrealistic. However, our estimators do not require estimating a joint distribution for $\varepsilon_{0T}$ and $\varepsilon_{1T}$, so dependence between $\varepsilon_{0T}$ and $\varepsilon_{1T}$ is irrelevant. This was verified in an additional set of simulations with $\varepsilon_{0T}$ and $\varepsilon_{1T}$ correlated (Figure A5 in the Supplemental Material).
We also conducted simulations under the scenario of null treatment effects by setting $\beta_3 = 0$. Our estimators performed well when the treatment did not impact the outcome with relatively low bias and coverage close to 95\% for all sample sizes considered (Supplemental Material Figure A6).\par

\section{Application}\label{Application}
We illustrate our methods in a two-group, two-period setting investigating the effect of Medicaid expansion on CD4 count at entry into clinical care among people living with HIV (PWH) in the United States. Medicaid expansion increased the number of people who were eligible to receive government-sponsored health care. A 2012 Supreme Court decision made Medicaid expansion optional. A total of 26 states decided to expand Medicaid in 2014. CD4 count is a continuous variable that is a marker of the strength of the immune system; CD4 decreases over time in PWH in the absence of treatment. Hence, a higher CD4 at entry into HIV care is desirable and reflects more timely HIV testing, diagnosis, and linkage to care.

We used data on 10,428 PWH from the North American AIDS Cohort Collaboration on Research and Design (NA-ACCORD).\cite{NA-ACCORD} These individuals were treatment naive (i.e., they had not received treatment for HIV prior to entering care). The value $T=0$ was assigned to PWH who enter HIV care in 2013 and $T=1$ was assigned to those entering HIV care in 2014. Individuals living in a state that did not expand Medicaid in 2014 were in the control group ($D=0$), while individuals living in a state that did expand in 2014 were in the treatment group ($D=1$). In our analysis, each individual only had one measurement (i.e., their CD4 count at entry into HIV care), and when and where they started HIV care assigned them to one of four groups defined by their values of $T$=\{0,1\} and $D$=\{0,1\}. We included as covariates race (Black vs. others), sex (male vs. female), age at entry into HIV care, and region (South vs. other).\cite{CensusRegions}

\begin{figure}[htbp]
    \begin{subfigure}[b]{\linewidth}
      \centering
	  \includegraphics[width = 12 cm]{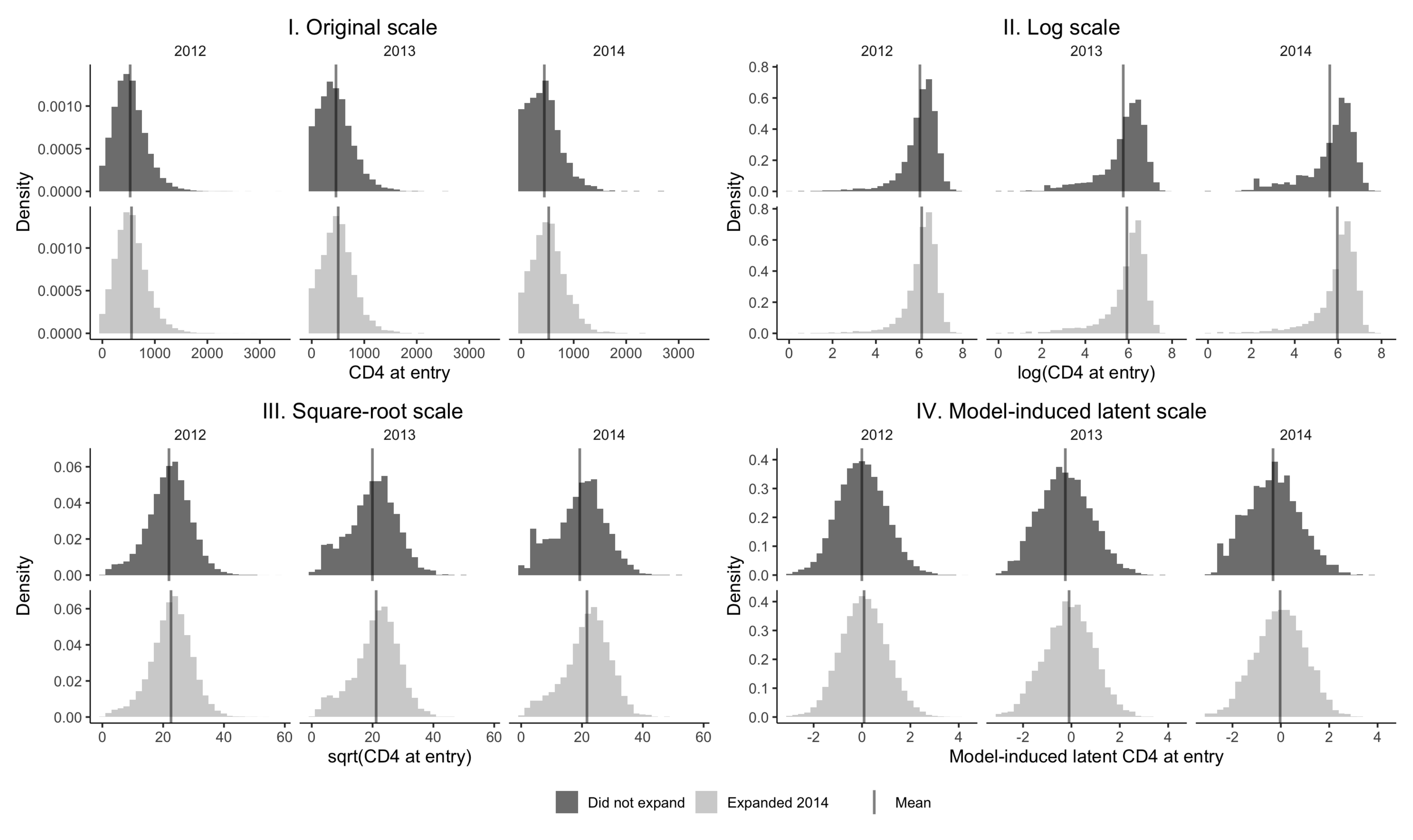}
    \end{subfigure}
    \vspace{0.5cm}
    \begin{subfigure}[b]{\linewidth}
	   \centering
	    \includegraphics[width = 12 cm]{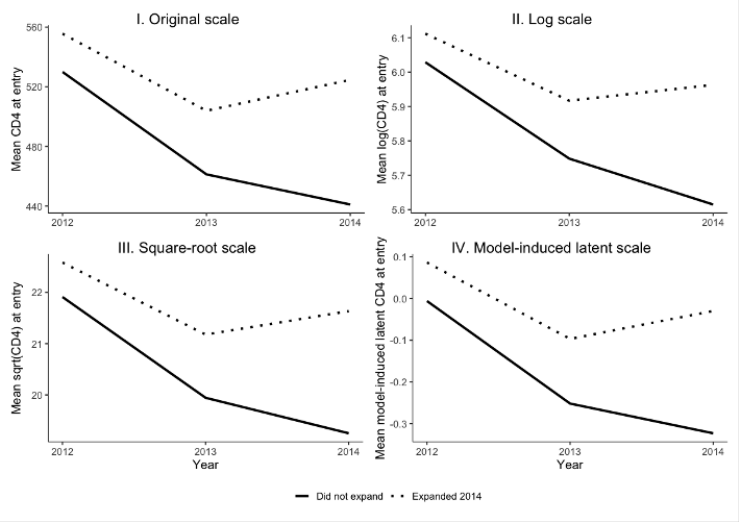}
    \end{subfigure}
    \caption{CD4 count at enrollment into care among PWH in 2012, 2013, and 2014, on the original scale, on the log scale, on the square-root scale, and on the cumulative probability model-induced latent scale. The upper panels display histograms, the bottom panels display means by calendar year.}
    \label{fig:y_hist}
\end{figure}

The distribution of CD4 count is right-skewed, and it is common to categorize or transform this outcome with a log or square-root transformation. Figure \ref{fig:y_hist} shows the distribution of CD4 count at enrollment in 2012, 2013, and 2014 in expansion states and non-expansion states on the original scale, after log-transformation, and after square-root transformation. \ref{fig:y_hist}  also displays the distribution on the CPM-induced latent scale, represented by the intercept values from an unadjusted version of model (\ref{cpm}) (i.e., $G\left[P\left(Y_{dt}\leq y|X\right)\right] = H^{-1}(y)$) with a probit link function. The traditional, unadjusted two-period, two-group DID with the outcome on the original scale relies on a parallel trends assumption on the original scale and thus results in an ATT interpreted on the original scale. Log- and square-root-transformed outcome data require parallel trends assumptions and ATT interpretation on the log- and square-root scales, respectively. Alternatively, our proposed approach requires the parallel trends assumption to hold on the mean of the model-induced latent scale while the ATT, and the other estimands, are interpreted on the original scale. Examining parallelism in pre-treatment years (2012 and 2013) can be used as a proxy, although with limitations, for assessing the reasonableness of the parallel trends assumption during the study period (2013 to 2014).\cite{Wing} Figure \ref{fig:y_hist} also shows the mean outcome on each scale in 2012, 2013, and 2014. While the parallel trends assumption appears reasonable in each scale (approximately parallel lines from 2012 to 2013), DID on the original scale relies on means of skewed data, while DID on the log and square-root scales complicates interpretation of the ATT. In addition, QTT, PTT, and MTT can also be estimated with our approach under the same parallel trends assumption used for estimating ATT.

We fit model (\ref{cpm}) to our data (including the covariates mentioned above) with a probit link function to estimate the ATT, $\text{QTT}(p)$ (for $p$ from 0.05 to 0.95 by 0.05), $\text{PTT}(y)$ (for $y$ from 50 to 750 by 50 cells/mm$^3$), and MTT. When summarizing results, we focus on the PTT for 200, 350, and 500 cells/mm$^3$ because these CD4 thresholds have historically been used in clinical practice and research.\par

Valid estimation requires Assumptions \ref{assump:cons}-\ref{assump:ltm} to hold. Assumption \ref{assump:cons} states that the observed CD4 for an individual is equal to the potential CD4 for that individual under their observed expansion status. Assumption \ref{assump:cons} could be violated if there are multiple versions of treatment, which actually happened in our study: all expansion states expanded Medicaid on January 1, 2014 except for Michigan and New Hampshire, which started later in 2014. However, fewer than 1\% of PWH in our study resided in these two states, so although there is some evidence of \ref{assump:cons} being violated, these violations are likely minor. Assumption \ref{assump:no_int} states that the potential CD4 for an individual is not affected by the expansion status of other individuals. In our context, this assumption could be violated if there was substantial interaction between study participants. Although there may be some interaction between participants at the same study sites, these are likely minor and do not meaningfully alter individual testing behavior. Assumption \ref{assump:cond_pta_prime} states that for individuals with the same set of covariates, the change over time in the mean CD4 on the latent scale for those in expansion states had there been no expansion would be the same as the change over time for those in non-expansion states. As mentioned earlier, \ref{assump:cond_pta_prime} is not testable; parallel trends assumptions on the original scale are similarly not testable. Figure \ref{fig:y_hist} (bottom right panel) suggests parallel trends might be reasonable, as in the absence of treatment from 2012 to 2013, the change in the expected CD4 on the latent scale was roughly the same between groups. However, the parallel trends assumption is made on 2013 to 2014 (not 2012 to 2013) in the absence of treatment which is counterfactual and unobservable. By conditioning on the covariates race, sex, age, and region, which may be associated with CD4 and imbalanced between expansion and non-expansion states, \ref{assump:cond_pta_prime} may be more plausible. With that said, \ref{assump:cond_pta_prime} may still not be valid as there are many other unmeasured factors that could have potentially altered temporal CD4 patterns between states (e.g., economic status). Finally, Assumption \ref{assump:ltm} implies that the CPM has been properly specified. This can be examined using residual diagnostics.\cite{liu_modeling_2017} Quantile-quantile (QQ) plots and residual-by-predictor plots suggest the probit link function is reasonable and that the predictors have been adequately included in the model. (See Figures A7 and A8 and Table A1 in the Supplemental Material for details.)\par

\begin{figure}[!ht]
	\centering
	\includegraphics[width = 12 cm]{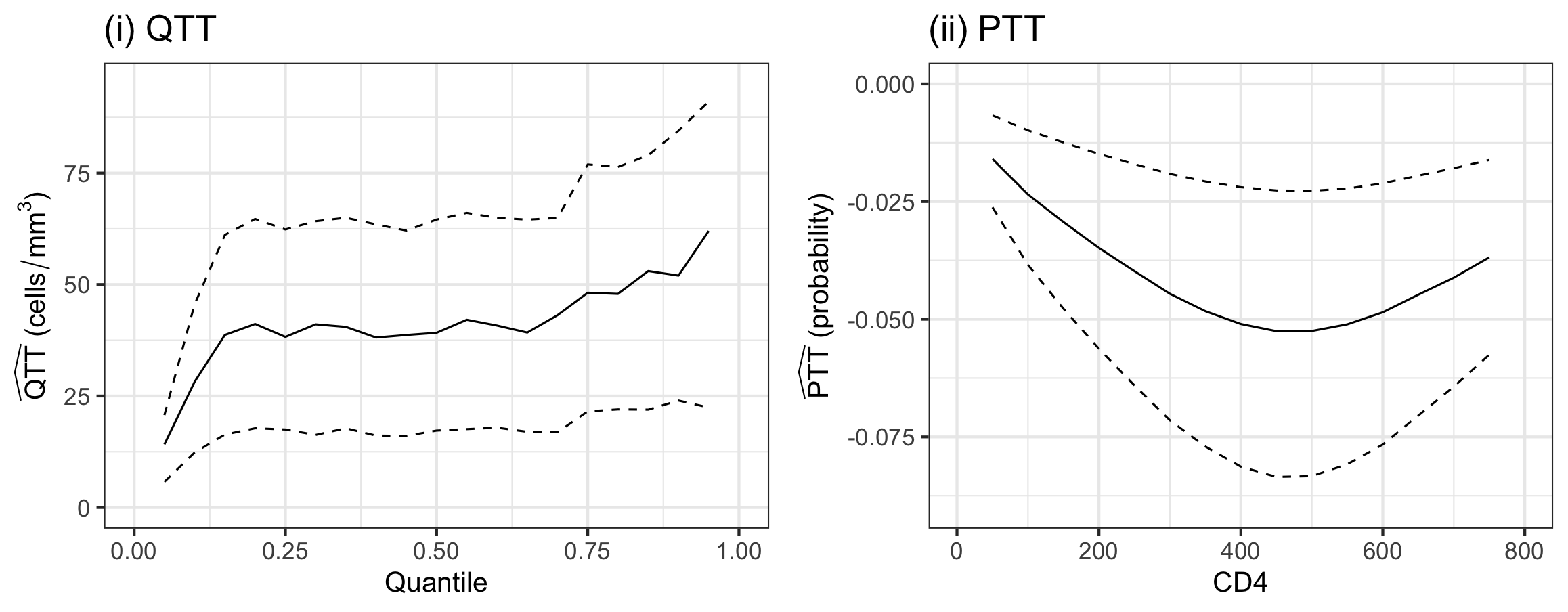}
		\caption{Point estimates and 95\% bootstrapped CIs (calculated as described in Section \ref{Inference}) of (i) QTT for different quantiles of the outcome and (ii) PTT for different values of the outcome.$^*$ \\\footnotesize{*QTT (quantile treatment effect among the treated), PTT (probability treatment effect among the treated).}}
		\label{fig:qtt_dtt_plot}
\end{figure}

Estimated treatment effects and bootstrapped 95\% CIs based on the fitted CPM are displayed in Table \ref{tab:probit_results}. The estimated ATT indicates Medicaid expansion in 2014 resulted in a higher CD4 count by an average of 41.1 cells/mm$^3$ (95\% CI of 17.7 to 68.1 cells/mm$^3$) among PWH in expansion states compared to if their state had not expanded Medicaid. The estimate of QTT$(0.50)$ is similar and indicates that expansion in 2014 resulted in median CD4 being 39.2 cells/mm$^3$ higher (95\% CI of 17.4 to 66.5 cells/mm$^3$) among PWH in expansion states compared to if their state had not expanded Medicaid. We see a slightly more pronounced impact at higher quantiles, e.g., QTT$(0.75)$. The estimated PTT$(500)$ indicates that the probability that CD4 among PWH was below 500 cells/mm$^3$ in expansion states reduced by 0.053 (95\% CI of $-$0.023 to $-$0.087) due to expansion, compared to no expansion. Lastly, the estimated MTT indicates that the probability that a randomly drawn individual in an expansion state would have a higher CD4 than another randomly drawn individual in an expansion state had there been no expansion is 0.537 (95\% CI of 0.516 to 0.562). This suggests that among PWH in expansion states, expansion resulted in higher CD4. Combined with all other QTT and PTT estimates (Figure \ref{fig:qtt_dtt_plot}), these results suggest that after adjusting for an individual's race, sex, age, and geographical region, the impact of Medicaid expansion on CD4 count at entry into HIV care in expansion states was beneficial.\par

\begin{table}
	\small\sf\centering
	\caption{Estimates and 95\% CIs (from 500 bootstrap iterations as described in Section~\ref{Inference}) with link function specified as probit for each treatment effect assessing the impact of Medicaid expansion on CD4 count at enrollment into care for PWH.}
	\label{tab:probit_results}
	\begin{tabular}{lcc}
		\hline
		Estimand & Estimate & Bootstrapped 95\% CI \\
		\hline
		ATT        & 41.1   & (17.7, 68.1) \\
		QTT(0.25)  & 38.3   & (16.5, 64.9) \\
		QTT(0.50)  & 39.2   & (17.4, 66.5) \\
		QTT(0.75)  & 48.2   & (21.6, 78.5) \\
		PTT(200)   & $-$0.035 & ($-$0.060, $-$0.015) \\
		PTT(350)   & $-$0.048 & ($-$0.081, $-$0.021) \\
		PTT(500)   & $-$0.053 & ($-$0.087, $-$0.023) \\
		MTT        & 0.537  & (0.516, 0.562) \\
		\hline
	\end{tabular}
\end{table}

To compare against our approach, we also estimated the ATT using the two other approaches described in Section \ref{Simulation}. The estimates, $\widehat{\text{ATT}}' = 35.6$ cells/mm$^3$ (traditional estimator; 95\% CI of 11.3 to 59.8 cells/mm$^3$), and $\widehat{\text{ATT}}'' = 83.4$ cells/mm$^3$ (Callaway \& Sant'Anna estimator \cite{Callaway}; 95\% CI of 40.9 to 125.8 cells/mm$^3$), also indicate a positive effect of Medicaid expansion on CD4. Alternatively, to avoid relying on means of skewed data, the estimated ATT using the approach of Callaway \& Sant'Anna \cite{Callaway} after square-root transforming the outcome was 2.2 $(\text{cells/mm}^3)^{\frac{1}{2}}$ (95\% CI of 1.1 to 3.3 $(\text{cells/mm}^3)^{\frac{1}{2}}$). These results are each based on a different parallel trends assumptions; the latter is on the square-root scale and estimates cannot be simply back-transformed for interpretation on the original scale.\par

We also compared our QTT estimates with an alternative approach that uses quantile DID modeling (\texttt{QDiD} function in the \texttt{qte} package\cite{qte}). The resulting estimates of QTT$(0.25)$, QTT$(0.50)$, and QTT$(0.75)$ using quantile regression were 95.8 (42.6 to 148.9), 84.2 (31.1 to 137.4), and  91.6 cells/mm$^3$ (32.2 to 151.0), respectively. These estimates using quantile regression are larger than our estimates, and they rely on different conditional parallel trends assumptions, a separate assumption for each quantile.\par

We then compared our PTT estimates with an alternative approach in which the outcome is dichotomized at a specified threshold and then a standard ATT approach is used to estimate the PTT at that threshold. For example, when the threshold is 200, we fit a linear model for $I\{y \leq 200\}$ on $D$, $T$, an interaction between $D$ and $T$, and all covariates; PTT$(200)$ is the estimated coefficient of the interaction term. The estimate of PTT$(200)$ was $-$0.037 (95\% CI of $-$0.068 to $-$0.005), of PTT$(350)$ was $-$0.041 (95\% CI of $-$0.078 to -0.004), and of PTT$(500)$ was $-$0.051 (95\% CI of $-$0.089 to $-$0.012). This approach makes unconditional parallel trends assumptions, a separate assumption for each threshold.\par 

Lastly, we compared our estimates with those from an alternative estimator that plugged in fitted values from distribution regression (DR) instead of CPMs, as described in Section \ref{Simulation}. Results using DR are displayed in Table A2. DR estimates were comparable to our estimates, although slightly closer to the null, for all treatment effects. Confidence intervals for the DR estimates tended to be wider, such that only those for ATT and QTT(0.25) excluded the null.\par

\section{Discussion}\label{Discussion}
In this paper we developed a distributional regression DID approach capable of estimating average, quantile, probability, and Mann-Whitney treatment effects among the treated with complicated, often difficult-to-model outcome distributions. To our knowledge, our approach is the first to provide estimation of an MTT. Our approach fits one semi-parametric CPM and makes a single conditional parallel trends assumption after an unspecified transformation of the outcome. Simulations indicate that our approach has good finite sample performance, even with relatively low sample sizes and highly skewed data. We showed how our approach can be used to provide a broad understanding of the impact of a policy using an illustration with Medicaid expansion.\par

Our methods are useful in classic DID settings, where an intervention is applied on a subset of subjects and data are available both pre- and post-intervention; the data may be longitudinal (subjects have data both pre- and post-intervention), cross-sectional (subjects only have pre- or post-intervention data), or a mixture of the two. Our methods are particularly designed to address settings with continuous outcomes that are skewed or subject to outliers. In these settings, we recommend presenting estimates of multiple estimands (i.e., the ATT, QTT, PTT, and MTT), thus providing a richer understanding of the effect of the intervention. We also recommend collecting as many variables as possible that may impact receiving the intervention and temporal patterns in the response variable, thus making the critical conditional parallel trends assumption more plausible.\par
Our proposed method requires a single parallel trends assumption conditional on covariates and is made on the means of an unspecified transformation of the outcome. This is appealing because we avoid comparing means of potentially skewed data without requiring a transformation of the outcome. This also avoids the need for separate parallel trends-type assumptions and the need for separate models, one for each estimand. Additionally, while we make a homogeneous treatment effect assumption on the latent scale by definition, our approach allows for heterogeneous treatment effects across covariate values on the original outcome scale.\par
Our approach is not without limitations. First, estimation can be computationally expensive with larger sample sizes, specifically when estimating the MTT, which involves pairwise comparisons of every subject; this could potentially be sped up via sampling. Second, as with all DID approaches, the parallel trends assumption is not directly testable, and it may be more challenging to conceptualize it on a latent scale. Third, our approach requires specifying a link function that defines the latent variable distribution; simulation results indicate our approach may provide biased estimates under link function misspecification. Lastly, although our method resulted in internally consistent estimates all derived with a single parallel trends assumption, some might reasonably argue that there are also benefits to performing analyses under a variety of assumptions to investigate robustness of findings.\par
Our current approach is designed for the two-group, two-period setting and future work could extend this to settings with more groups and more time periods.\cite{Callaway} While we focused on skewed continuous outcomes, given that CPMs can be fit to many outcome types, investigations of our approach to estimate the QTT, PTT, and MTT in other settings such as those with ordinal outcomes or continuous outcomes with detection limits are warranted.\par 

\clearpage
\bibliographystyle{SageV}
\bibliography{mybibilo}
\clearpage
\setcounter{figure}{0}
\renewcommand{\thefigure}{A\arabic{figure}}
\setcounter{table}{0}
\renewcommand{\thetable}{A\arabic{table}}
\setstretch{1}

\section*{Supplementary Material}

\subsection*{Proof of Identification Result 2}

First note that $Y^*_{0|11} = \text{E}\left[Y_{0|11}^*|X\right] + \varepsilon - c$, where $c = \text{E}[\varepsilon$] and $\varepsilon \sim F_{\varepsilon}$. Thus, $P\left(Y_{0|11} \leq y |X\right) = F_{\varepsilon}\left(H^{-1}(y) - \text{E}\left[Y_{0|11}^* | X\right] + c\right)$. Therefore,
\begin{flalign*}
F_{Y_{0|11}}(y) &= P\left(Y_{0|11} \leq y\right)\\
&= \int P\left(Y_{0|11} \leq y |X = x\right)dF(x|D=1,T=1)\\
&= \int F_{\varepsilon}\left(H^{-1}(y) - \left(\text{E}[Y^{*}_{0|01}|X = x] - \text{E}[Y^{*}_{0|00}|X = x] + \text{E}[Y^{*}_{0|10}|X = x]\right) + c\right)dF(x|D=1,T=1)&\\
&= \int F_{\varepsilon}\left(H^{-1}(y) - \text{E}[Y_{01}^{*}|X = x] + \text{E}[Y_{00}^{*}|X = x] - \text{E}[Y_{10}^{*}|X = x] + c\right)dF(x|D=1,T=1)&\\
&= \int F_{\varepsilon}\left(H^{-1}(y) - (\beta_2 + \beta_4^T x + c) + \beta_4^T x + c -(\beta_1 + \beta_4^Tx + c) + c\right)dF(x|D=1,T=1)&\\
&= \int F_{\varepsilon}\left(H^{-1}(y) - \beta_1 - \beta_2 - \beta_4^Tx\right)dF(x|D=1,T=1),&
\end{flalign*}

\noindent where the first line follows from A3', the second follows from A1, and the third line follows from A4.\par

\begin{figure}[h!]
	\centering
	\includegraphics[width = 17.5 cm]{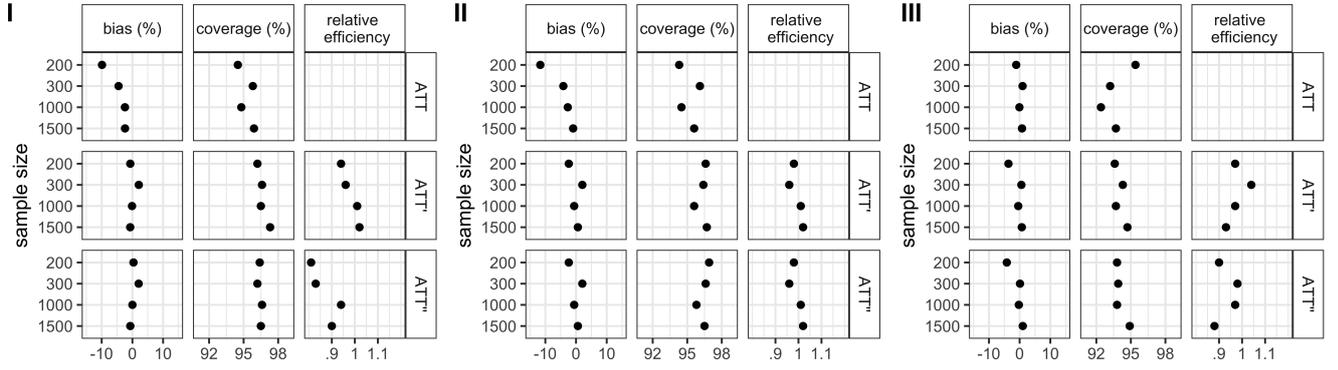}
		\begin{singlespace}
	\caption{Percent bias, coverage, and relative efficiency of ATT estimators from simulation with 1000 replications with 200 bootstrap iterations (for $\widehat{\text{ATT}}$ coverage calculation only as described in Section 4.2) for the scenario in which \textbf{I}: $Y = D + 0.5T + 0.5DT + 0.25X_1 + 0.5X_2 + \varepsilon_T$, and $\varepsilon_T \sim \text{Normal}(0,1)$, with link function specified as probit; \textbf{II}: $Y = D + 0.5T + 0.5DT + \varepsilon_T$, and $\varepsilon_T \sim \text{Normal}(0,1)$, with link function specified as probit; and \textbf{III}: $Y = D + 0.5T + 0.5DT + 0.25X_1 + 0.5X_2 + \varepsilon_T$, and $\varepsilon_T \sim \text{Logistic}(0,1)$, with link function specified as logit. Relative efficiency calculated as $MSE_{\widehat{\text{ATT}}}/MSE_{\widehat{\text{ATT}}'}$ and $MSE_{\widehat{\text{ATT}}}/MSE_{\widehat{\text{ATT}}''}$, where MSE denotes mean squared error. The true value is $\text{ATT} = 0.50$.$^*$ \\\footnotesize{*ATT (average treatment effect among the treated), MSE (mean squared error).}}
		\label{fig:scenario3_7_8}
		\end{singlespace}

\end{figure}

\clearpage\begin{figure}[h!]
	\centering
	\includegraphics[width = 5.5 cm]{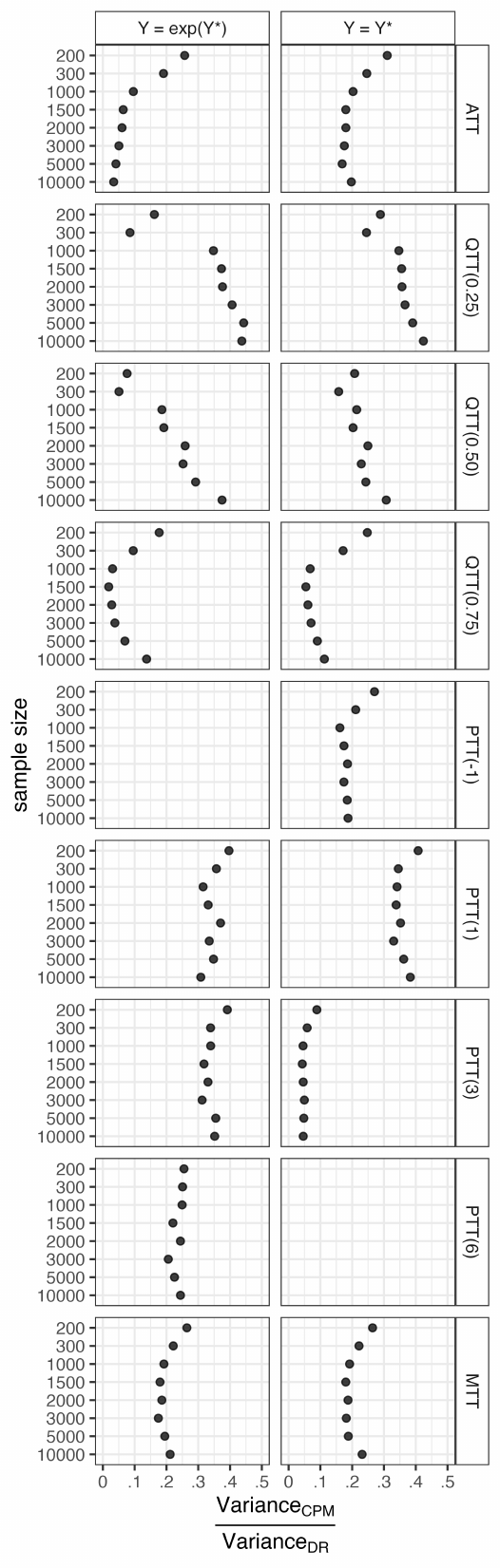}
	\begin{singlespace}
		\caption{Relative variance of ATT, QTT, PTT, and MTT estimators using the CPM approach and the DR approach from simulation with 1000 replications for the scenario in which $Y = exp(D + 0.5T + 0.5DT + 0.25X_1 + 0.5X_2 + \varepsilon_T)$ and $Y = D + 0.5T + 0.5DT + 0.25X_1 + 0.5X_2 + \varepsilon_T$, with $\varepsilon_T \sim \text{Normal}(0,1)$ and link function specified as probit for both scenarios. Relative variance for the ATT is calculated as $\text{Variance}(\widehat{\text{ATT}}_{CPM})/\text{Variance}(\widehat{\text{ATT}}_{DR})$. Relative variance for all other estimators are calculated with corresponding variance values. True values for $Y = exp(Y^*)$ are as stated in Section 5. True values for $Y = Y^*$ are ATT = QTT($\cdot$) = 0.50, PTT($-1$) = $-$0.007, PTT($1$) = $-$0.13, PTT($3$) = $-$0.11, and MTT = 0.623.$^*$ \\\footnotesize{*ATT (average treatment effect among the treated), QTT (quantile treatment effect among the treated), PTT (probability treatment effect among the treated), MTT (Mann-Whitney treatment effect among the treated), CPM (cumulative probability model), DR (distributional regression).}}
			\label{fig:scenario15_35_RE}
	\end{singlespace}
\end{figure}

\clearpage\begin{figure}[h!]
	\centering
	\includegraphics[width = 12 cm]{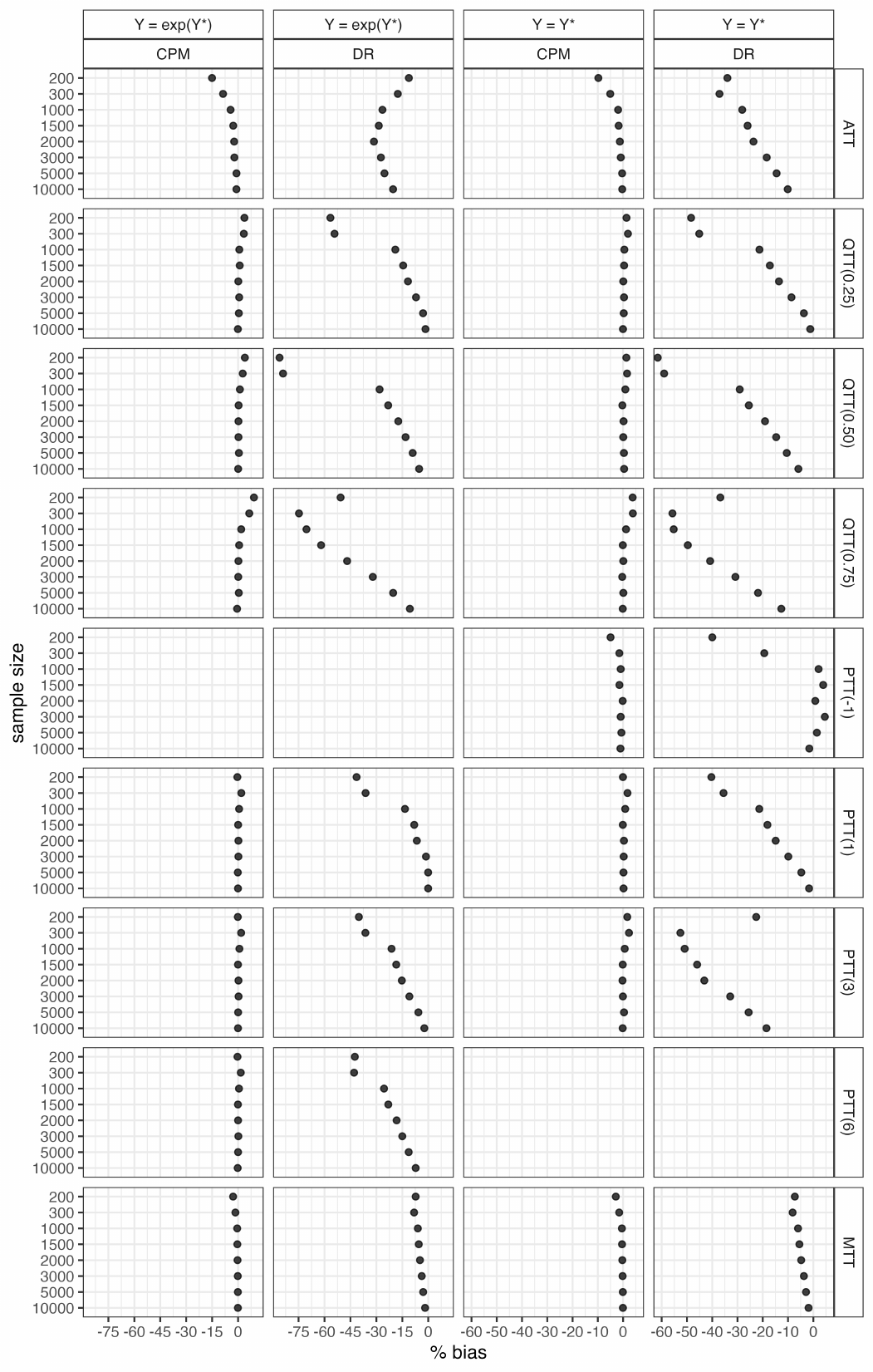}
	\begin{singlespace}
		\caption{Percent bias of ATT, QTT, PTT, and MTT estimators using the CPM approach and the DR approach from simulation with 1000 replications for the scenario in which $Y = exp(D + 0.5T + 0.5DT + 0.25X_1 + 0.5X_2 + \varepsilon_T)$ and $Y = D + 0.5T + 0.5DT + 0.25X_1 + 0.5X_2 + \varepsilon_T$, with $\varepsilon_T \sim \text{Normal}(0,1)$ and link function specified as probit for both scenarios. True values for $Y = exp(Y^*)$ are as stated in Section 5. True values for $Y = Y^*$ are ATT = QTT($\cdot$) = 0.50, PTT($-1$) = $-$0.007, PTT($1$) = $-$0.13, PTT($3$) = $-$0.11, and MTT = 0.623.$^*$ \\\footnotesize{*ATT (average treatment effect among the treated), QTT (quantile treatment effect among the treated), PTT (probability treatment effect among the treated), MTT (Mann-Whitney treatment effect among the treated), CPM (cumulative probability model), DR (distributional regression).}}
		\label{fig:scenario15_35_PB}
	\end{singlespace}
\end{figure}

\clearpage
\begin{figure}[h!]
	\centering
	\includegraphics[width = 11 cm]{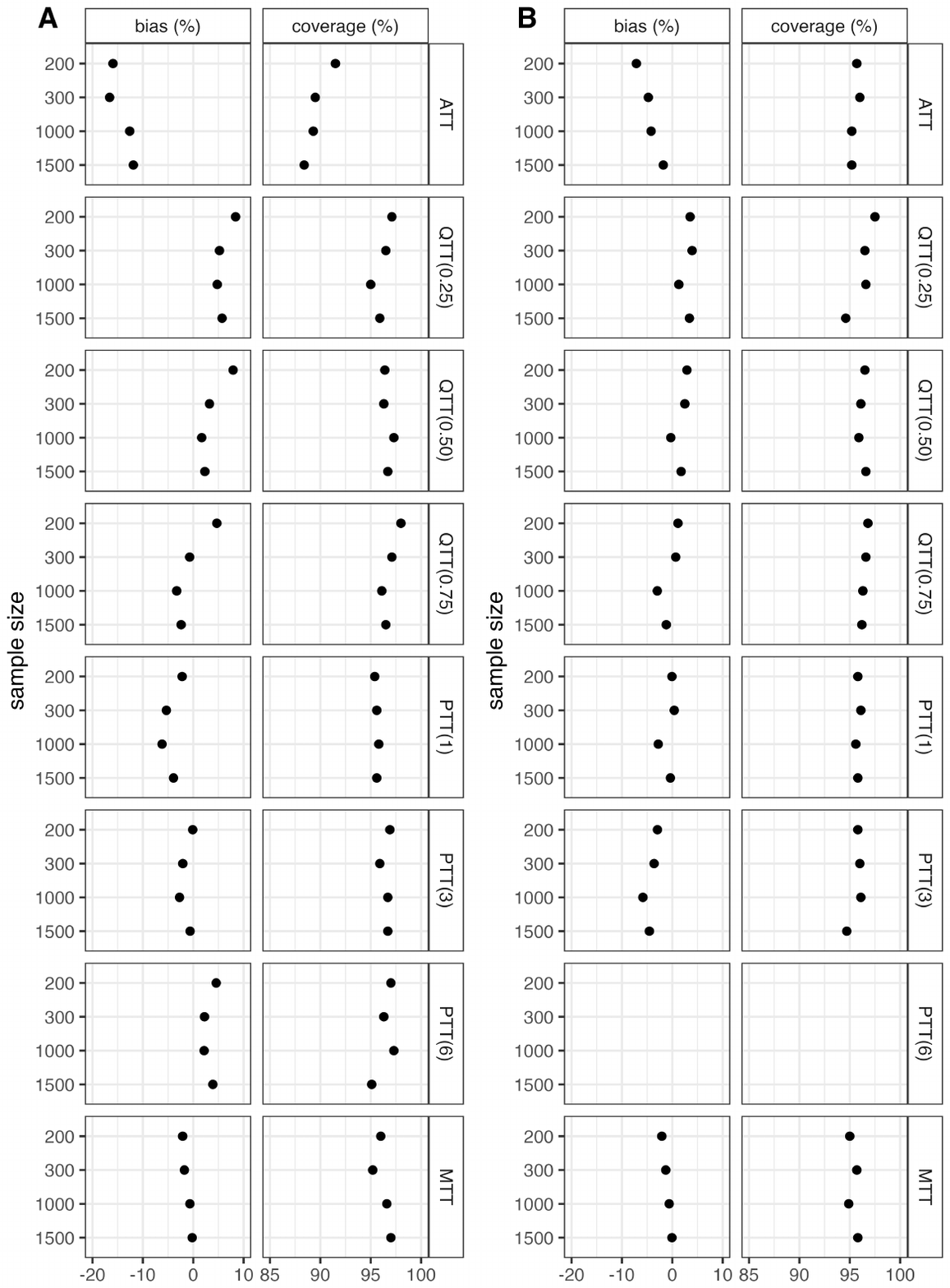}
		\begin{singlespace}
	\caption{Percent bias and coverage of ATT, QTT$(0.25)$, QTT$(0.50)$, QTT$(0.75)$, PTT$(1)$, PTT$(3)$, PTT$(6)$ (panel \textbf{A} only), and MTT estimators from simulation of 1000 replications with 200 bootstrap iterations (as described in Section 4.2) for the scenario in which \textbf{A}: $Y = \exp(D + 0.5T + 0.5DT + 0.25X_1 + 0.5X_2 + \varepsilon_T)$, and $\varepsilon_T \sim \text{Normal}(0,1)$, with link function misspecified as logit; \textbf{B}: $Y = D + 0.5T + 0.5DT + 0.25X_1 + 0.5X_2 + \varepsilon_T$, and $\varepsilon_T \sim \text{Normal}(0,1)$, with link function misspecified as logit. The true values for \textbf{A} are as stated in Section 5. The true values for \textbf{B} are ATT = QTT($\cdot$) = 0.50, PTT($1$) = $-$0.13, PTT($3$) = $-$0.11, and MTT = 0.623.$^*$. PTT$(6)$ was not calculated for scenario \textbf{B} because the maximum value of $Y$ when $Y = D + 0.5T + 0.5DT + 0.25X_1 + 0.5X_2 + \varepsilon_T$, and $\varepsilon_T \sim \text{Normal}(0,1)$ is less than $6$. \\\footnotesize{*ATT (average treatment effect among the treated), QTT (quantile treatment effect among the treated), PTT (probability treatment effect among the treated), MTT (Mann-Whitney treatment effect among the treated).}}
		\label{fig:scenario2_4}
	\end{singlespace}

\end{figure}

\clearpage
\begin{figure}[h!]
	\centering
	\includegraphics[width = 5.5 cm]{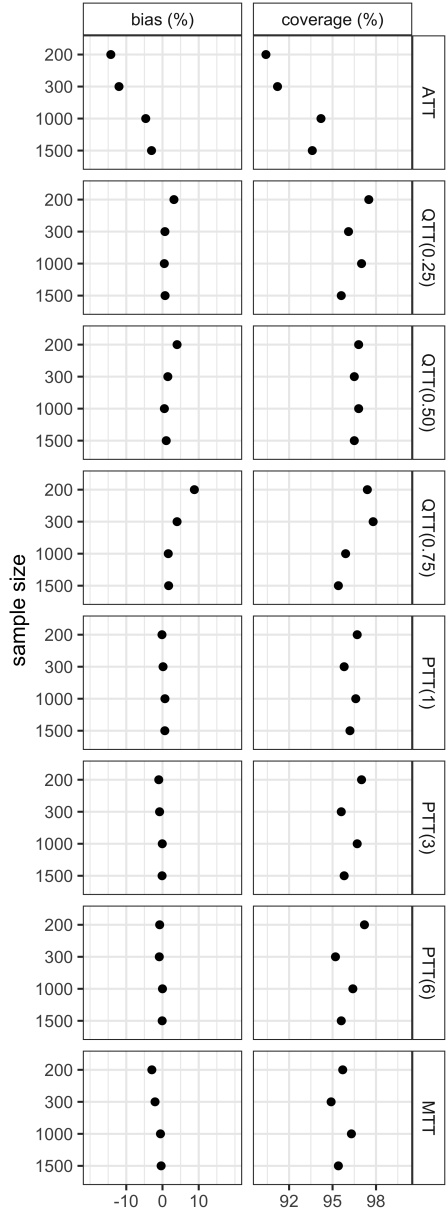}
	\begin{singlespace}
\caption{Percent bias and coverage of ATT, QTT$(0.25)$, QTT$(0.50)$, QTT$(0.75)$, PTT$(1)$, PTT$(3)$, PTT$(6)$, and MTT estimators from simulation of 1000 replications with 200 bootstrap iterations (as described in Section 4.2) for the scenario in which $Y = \exp(D + 0.5T + 0.25X_1 + 0.5X_2 + \varepsilon_T)$ with link function specified as probit. For those with observations at time $T = 0$ and $T=1$, we generated $\left(\varepsilon_{00}, \varepsilon_{01}, \varepsilon_{10}, \varepsilon_{11}\right)$ from $\text{Normal}\left(\protect\begin{pmatrix}0 \\ 0 \\ 0 \\ 0\protect\end{pmatrix}, \protect\begin{pmatrix}
	1 & 0.5 & 0.5 & 0.5 \\ 
	0.5 & 1 & 0.5 & 0.5 \\ 
	0.5 & 0.5 & 1 & 0.5 \\ 
	0.5 & 0.5 & 0.5 & 1
	\protect\end{pmatrix}\right).$ For those with one observation at time $T = 0$ or $T=1$, we generated $\left(\varepsilon_{1T}, \varepsilon_{0T}\right)$ from $\text{Normal}\left(\protect\begin{pmatrix}0 \\ 0 \protect\end{pmatrix}, \protect\begin{pmatrix}
	1 & 0.5  \\ 0.5 & 1
	\protect\end{pmatrix}\right).$ The true values are as stated in Section 5.$^*$ \\\footnotesize{*ATT (average treatment effect among the treated), QTT (quantile treatment effect among the treated), PTT (probability treatment effect among the treated), MTT (Mann-Whitney treatment effect among the treated).}}
		\label{fig:scenario5}
	\end{singlespace}

\end{figure}

\clearpage
\begin{figure}[h!]
	\centering
	\includegraphics[width = 6 cm]{web_figure6.pdf}
	\begin{singlespace}
	\caption{Bias and coverage of ATT, QTT$(0.25)$, QTT$(0.50)$, QTT$(0.75)$, PTT$(1)$, PTT$(3)$, PTT$(6)$, and MTT estimators from simulation of 1000 replications with 200 bootstrap iterations (as described in Section 4.2) for the scenario in which $Y = \exp(D + 0.5T + 0.25X_1 + 0.5X_2 + \varepsilon_T)$ (i.e., null treatment effect), and $\varepsilon_T \sim \text{Normal}(0,1)$, with link function specified as probit.  The true values are null effects of $\text{ATT} = \text{QTT}(\cdot) = \text{PTT}(\cdot) = 0$ and $\text{MTT} = 0.50$.$^*$ \\\footnotesize{*ATT (average treatment effect among the treated), QTT (quantile treatment effect among the treated), PTT (probability treatment effect among the treated), MTT (Mann-Whitney treatment effect among the treated).}}
		\label{fig:scenario6}
	\end{singlespace}

\end{figure}

\clearpage
\begin{figure}[h!]
	\centering
	\includegraphics[width = 12 cm]{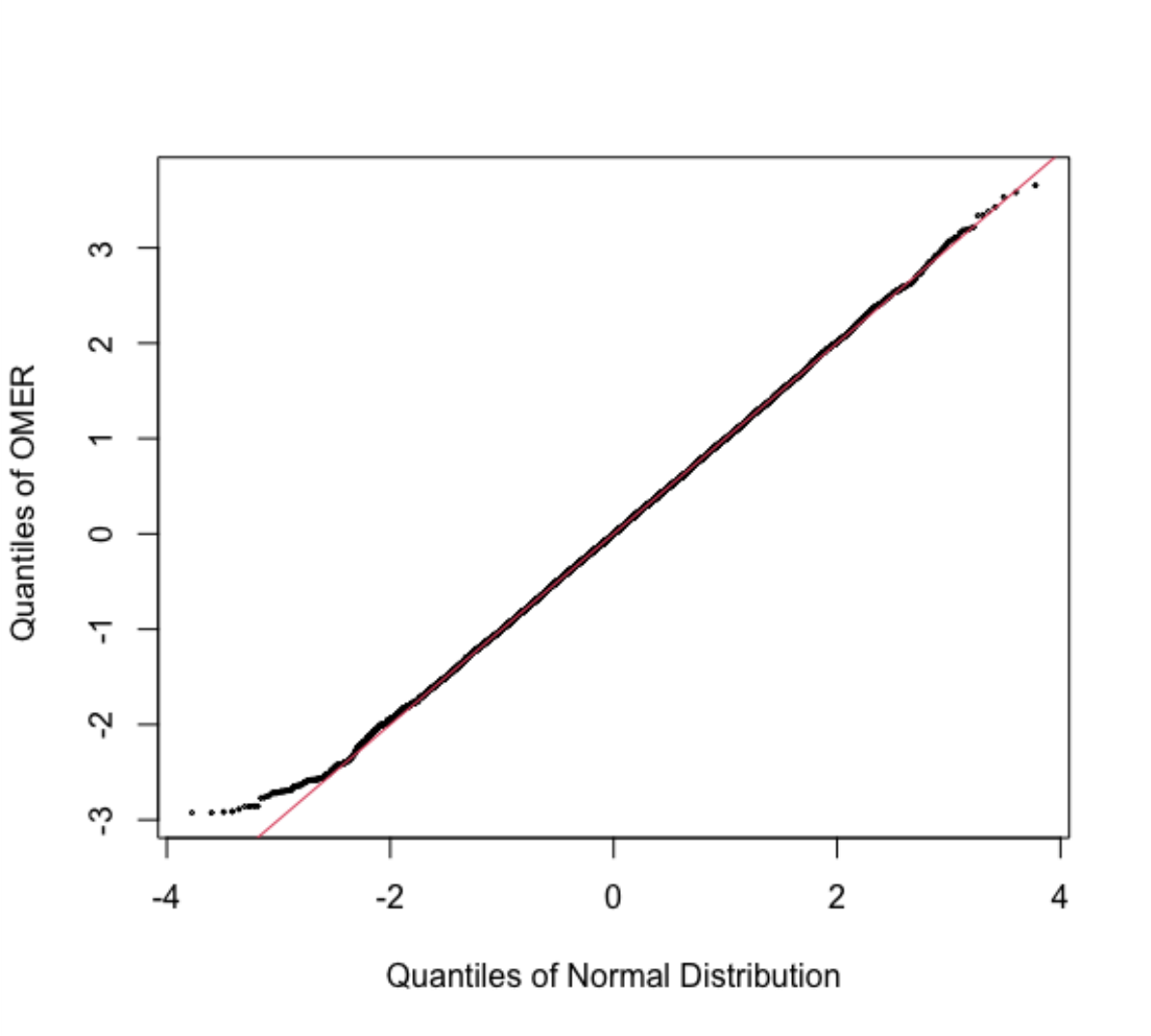}
		\begin{singlespace}
	\caption{QQ-plot of observed-minus-expected residuals (OMERs) for CPM described in Section 6.$^*$ \\\footnotesize{*CPM (cumulative probability model).}}
		\label{fig:qqplot}
	\end{singlespace}

\end{figure}

\clearpage
\begin{figure}[h!]
	\centering
	\includegraphics[width = 12 cm]{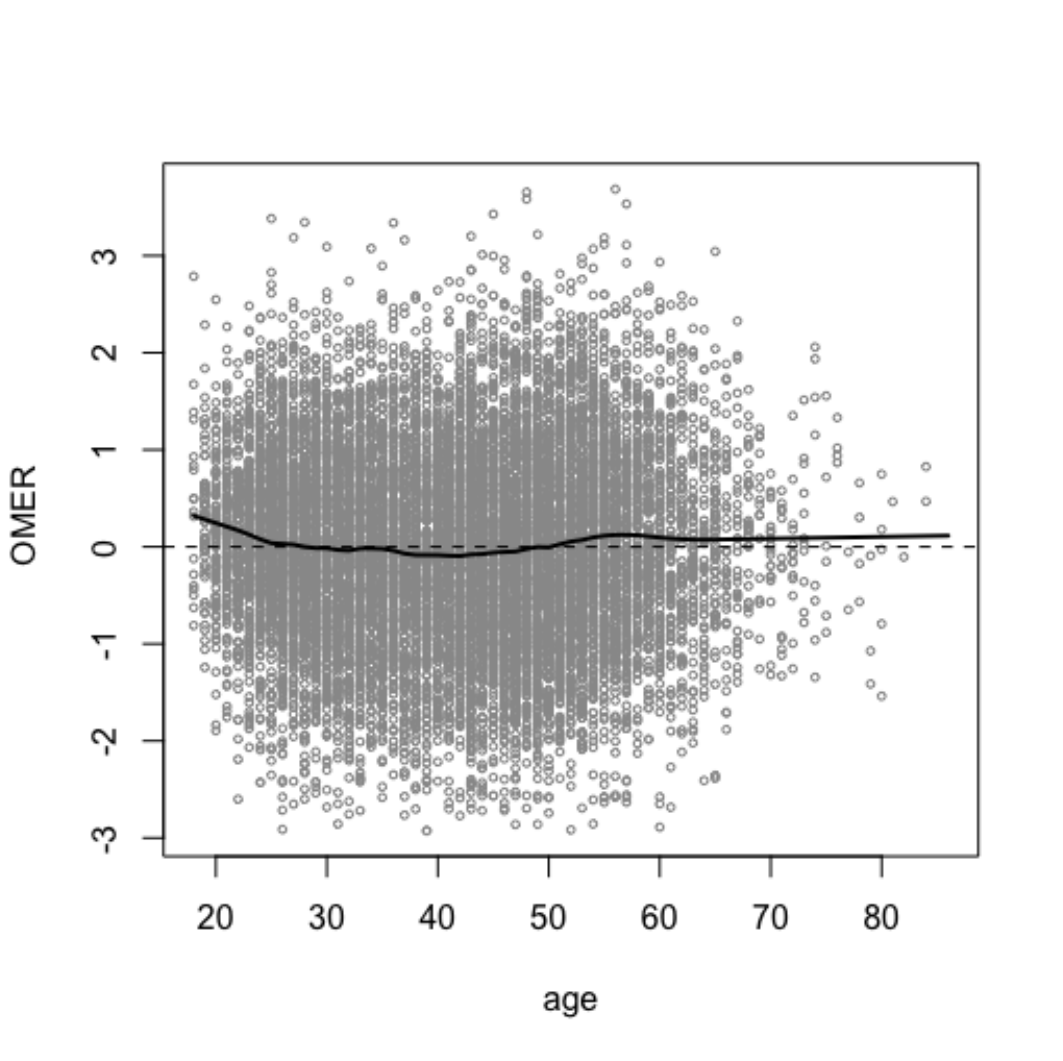}
		\begin{singlespace}
	\caption{Residual-by-predictor plots for age using observed-minus-expected residuals (OMERs) from CPM described in Section 6.$^*$ \\\footnotesize{*CPM (cumulative probability model).}}
		\label{fig:residplot}
	\end{singlespace}
\end{figure}

\clearpage
\begingroup
\renewcommand{\arraystretch}{0.90}
\begin{table}[h!]
		\caption{Estimates and 95\% CIs (from 500 bootstrap iterations as described in Section 4.2) with link function specified as cloglog assessing the impact of Medicaid expansion on CD4 count at enrollment into care for PWH.$^*$ \\\footnotesize{*ATT (average treatment effect among the treated), QTT (quantile treatment effect among the treated), PTT (probability treatment effect among the treated), MTT (Mann-Whitney treatment effect among the treated), PWH (people living with HIV).}\label{tab:cloglog_results}}
	\centering

	\begin{tabular}{lcc}
		\hhline{===}
		& Estimate & Bootstrapped 95\% CI \\
		\hhline{===}
		ATT & 29.1&  (14.2, 44.3)\\
		
		QTT(0.25) & 11.2 & (5.5, 19.0)  \\
		
		QTT(0.50) & 45.0 & (21.1, 70.2) \\
		
		QTT(0.75) & 36.9 &  (17.1, 57.0)\\
		
		PTT(200) & $-$0.046 &  ($-$0.071, $-$0.022)\\
		
		PTT(350) & $-$0.052 &($-$0.078, $-$0.023)  \\
		
		PTT(500) & $-$0.045 & ($-$0.068, $-$0.022)\\
		
		MTT & 0.535 &  (0.517, 0.553) \\
		\hline
	\end{tabular}

\end{table}

As a sensitivity analysis, we also repeated the example analysis in Section 6 with a cloglog link function. The likelihood for the fitted CPM with the cloglog link function was lower than that with the probit link function ($-$87137 vs. $-$87133), which suggests the probit is a better model fit. Details and results are in Table \ref{tab:cloglog_results}. In short, although we see slight differences in the magnitudes of the treatment effects with different link functions, the direction of all estimates are the same for both link functions considered and suggest an overall positive impact of  Medicaid expansion on CD4 count at enrollment among PWH living in expansion states.\par
\endgroup

\clearpage
\begingroup
\renewcommand{\arraystretch}{0.90}
\begin{table}[h!]
	\caption{Estimates and 95\% CIs (from 500 bootstrap iterations as described in Section 4.2) using a distribution regression method with link function specified as probit assessing the impact of Medicaid expansion on CD4 count at enrollment into care for PWH.$^*$ \\\footnotesize{*ATT (average treatment effect among the treated), QTT (quantile treatment effect among the treated), PTT (probability treatment effect among the treated), MTT (Mann-Whitney treatment effect among the treated), PWH (people living with HIV).}\label{tab:dr_results}}
	\centering
	
	\begin{tabular}{lcc}
		\hhline{===}
		& Estimate & Bootstrapped 95\% CI \\
		\hhline{===}
		ATT & 34.2&  (8.8, 57.3)\\
		
		QTT(0.25) & 38.7 & (9.2, 74.0)  \\
		
		QTT(0.50) & 35.0 & (0.20, 69.9) \\
		
		QTT(0.75) & 26.1 &  (-1.0, 85.5)\\
		
		PTT(200) & $-$0.042 &  ($-$0.082, $-$0.004)\\
		
		PTT(350) & $-$0.043 &($-$0.082, $-$0.002 )  \\
		
		PTT(500) & $-$0.048 & ($-$0.079, $-$0.008)\\
		
		MTT & 0.531 &  (0.506, 0.555) \\
		\hline
	\end{tabular}
	
\end{table}
\endgroup

\end{document}